\documentclass[preprint]{aastex}
\usepackage{color}

\shorttitle{A Multiple System Orbiting HD 1160}
\shortauthors{Nielsen et al.}

\begin{document}

\title{The Gemini NICI Planet-Finding Campaign: Discovery of a 
Multiple System Orbiting the Young A Star HD 1160}

\author{Eric L. Nielsen,\altaffilmark{1}
Michael C. Liu,\altaffilmark{1}
Zahed Wahhaj,\altaffilmark{1}
Beth A. Biller,\altaffilmark{2}
Thomas L. Hayward,\altaffilmark{3}
Alan Boss,\altaffilmark{4}
Brendan Bowler,\altaffilmark{1}
Adam Kraus,\altaffilmark{1,5}
Evgenya L. Shkolnik,\altaffilmark{6}
Matthias Tecza,\altaffilmark{7}
Mark Chun,\altaffilmark{1}
Fraser Clarke,\altaffilmark{7}
Laird M. Close,\altaffilmark{8}
Christ Ftaclas,\altaffilmark{1}
Markus Hartung,\altaffilmark{8}
Jared R. Males,\altaffilmark{8}
I. Neill Reid,\altaffilmark{9}
Andrew J. Skemer,\altaffilmark{8}
Silvia H. P. Alencar,\altaffilmark{10}
Adam Burrows,\altaffilmark{11}
Elisabethe de Gouveia Dal Pino,\altaffilmark{12}
Jane Gregorio-Hetem,\altaffilmark{12}
Marc Kuchner,\altaffilmark{13}
Niranjan Thatte,\altaffilmark{6}
Douglas W. Toomey\altaffilmark{14}
}

\altaffiltext{1}{Institute for Astronomy, University of Hawaii, 2680 
Woodlawn Drive, Honolulu HI 96822, USA}
\altaffiltext{2}{Max-Planck-Institut f\"ur Astronomie, K\"onigstuhl 17, 
69117 Heidelberg, Germany}
\altaffiltext{3}{Gemini Observatory, Southern Operations Center, c/o AURA,
Casilla 603, La Serena, Chile}
\altaffiltext{4}{Department of Terrestrial Magnetism, Carnegie Institution of 
Washington, 5241 Broad Branch Road, N.W., Washington, DC 20015, USA}
\altaffiltext{5}{Hubble Fellow}
\altaffiltext{6}{Lowell Observatory, 1400 West Mars Road, Flagstaff, AZ 
86001, USA}
\altaffiltext{7}{Department of Astronomy, University of Oxford, DWB, Keble 
Road, Oxford OX1 3RH, UK}
\altaffiltext{8}{Steward Observatory, University of Arizona, 933 North Cherry 
Avenue, Tucson, AZ 85721, USA}
\altaffiltext{9}{Space Telescope Science Institute, 3700 San Martin Drive, 
Baltimore, MD 21218, USA}
\altaffiltext{10}{Departamento de Fisica - ICEx - Universidade Federal de 
Minas Gerais, Av. Antonio Carlos, 6627, 30270-901, Belo Horizonte, MG, 
Brazil}
\altaffiltext{11}{Department of Astrophysical Sciences, Peyton Hall, 
Princeton University, Pinrceton, NJ 08544}
\altaffiltext{12}{Universidade de Sao Paulo, IAG/USP, Departamento de 
Astronomia, Rua do Matao, 1226, 05508-900, Sao Paulo, SP, Brazil}
\altaffiltext{13}{NASA Goddard Space Flight Center, Exoplanets and Stellar 
Astrophyics Laboratory, Greenbelt, MD, 20771, USA}
\altaffiltext{14}{Mauna Kea Infrared, LLC, 21 Pookela St., Hilo, HI 
96720, USA}

\begin{abstract}
We report the discovery by the Gemini NICI Planet-Finding Campaign of two 
low-mass companions to the young A0V star HD 1160 at projected separations of 
81 $\pm$ 5 AU (HD~1160~B) and 533 $\pm$ 25 AU (HD~1160~C).  VLT images 
of the system taken over a decade for the purpose of using HD~1160~A as a 
photometric calibrator confirm that both companions are physically 
associated.  By comparing the system to members of young moving groups and 
open clusters with well-established ages, we estimate an age of 
50$^{+50}_{-40}$ Myr for HD~1160~ABC.  While the 
$UVW$ motion of the system does not match any known moving group, the 
small magnitude of the space velocity is consistent with youth.  
Near-IR spectroscopy shows HD~1160~C to be an M3.5 $\pm$ 0.5 
star with an estimated mass of 0.22$^{+0.03}_{-0.04}$~M$_{\odot}$, 
while NIR photometry of HD~1160~B suggests a brown dwarf with a mass of 
33$^{+12}_{-9}$ M$_{Jup}$.  The very small mass ratio (0.014) 
between the A and B components of the system is rare for A star binaries, and 
would represent a planetary-mass companion were HD 1160 A to be slightly 
less massive than the Sun.
\end{abstract}


\section{Introduction}

There are almost 700 extrasolar planets discovered to date, most of them 
at small separations ($<$5 AU) from their parent stars 
\citep[e.g.][]{rvref,marcy08,mayor09}.  In addition, 1235 transiting planet 
candidates have been discovered by NASA's $Kepler$ Mission 
\citep{kepler11}, effectively tripling the census of exoplanet candidates.  
Unlike radial velocity and transit surveys, which are responsible 
for the vast majority of these discoveries, direct 
imaging allows us to access planets at much larger 
separations ($\gtrsim$ 5 AU).  Direct imaging surveys 
for gas-giant planets around young stars initially returned null 
results \citep{elena,gdps,sdifinal,sdifinalpab}, and a detailed analysis of 
these 
null results shows that giant planets are rare at large ($\gtrsim$~65~AU) 
separations around solar-type stars \citep{sdimeta}.  Despite the rarity of 
these objects, discoveries of directly imaged 
planets at large separations have now been announced 
\citep{hr8799b,hr8799e,fomalhautb,rxs1609_1,betapicb,lkca15b}, 
allowing the study of the 
composition and formation of this new population of planets 
\citep[e.g.][]{bowler_8799,currie11,barman11}.

High-mass stars have recently become exciting targets for direct imaging 
surveys.  Both brown dwarfs 
(e.g. HR 7329 B [\citealt{lowrance00}] and HR 6037 B [\citealt{hr6037b}]) and 
planetary mass companions (HR 8799 bcde, Fomalhaut b, and Beta Pic b 
[\citealp{hr8799e,fomalhautb,betapicb}]) 
have been detected around A-type stars.  
Analysis of radial-velocity results by \citet{johnson_new} has shown 
that short-period ($a$~$<$~2.5 AU) giant planets
are more frequent around high-mass A stars 
than low-mass M stars, and the direct detection of planets around 
multiple A stars (where the higher contrast ratio should make planet 
detection by direct imaging more difficult) suggests this trend may 
continue to larger separations \citep[e.g.][]{crepp11}.

The Gemini NICI Planet-Finding Campaign is a 3-year survey begun in 
December 2008 targeting about 300 young, nearby stars to directly image 
extrasolar giant planets \citep{liunici}.  
The goals of the Campaign are to detect new planets and to investigate 
the properties and formation mechanism of planets at $\gtrsim$10 AU 
separations.  
Designed specifically to detect planets, the NICI instrument 
\citep{toomey03,chun08} consists of an 
85-element curvature adaptive optics system, a Lyot coronagraph, and two 
cameras that can image simultaneously for spectral difference 
imaging \citep[SDI;][]{racinesdi}, observing through two moderate-band filters 
around the $H$-band methane feature seen in cool substellar objects 
\citep{burrows97,baraffe03}.  
NICI also uses angular differential imaging (ADI), which 
holds the telescope rotator fixed, allowing speckle noise to be 
separated from real companions \citep{liuadi,adi}.

We previously announced the Campaign discovery of two brown dwarfs orbiting 
young stars PZ Tel \citep{pztel} and CD-35 2722 \citep{cd35}.  We 
present here the discovery of two low-mass companions to the young 
A star HD 1160, at $\sim$80 AU (B) and $\sim$530 AU (C) separations.  
HD 1160 (see Table \ref{table1}) is a 
well known photometric standard star from \citet{elias82} and was selected 
for the Campaign due to its underluminous 
position on the HR diagram, a sign of youth for field A stars (e.g. 
\citealt{jura98}).  HD~1160 was observed by \citet{su06} with $Spitzer$ at 
24 $\mu$m, and no excess was detected.  We derive an age range 
for the system of 50$^{+50}_{-40}$ Myr and estimate masses for the 
two companions of 33$^{+12}_{-9}$ M$_{Jup}$ (HD~1160~B) 
and 0.22$^{+0.03}_{-0.04}$ M$_{\odot}$ (HD~1160~C).

\section{Observations}

\subsection{Gemini-South NICI: Near-Infrared Imaging}

NICI, the Near Infrared Coronagraphic Imager, is a dual-channel high contrast 
imager at Gemini South, designed specifically to directly image giant 
planets.  The detectors in the two channels are 1024x1024 Aladdin II InSb, 
with a square 18.43'' field of view and 17.94 and 17.96 mas pixels in the 
two cameras.

HD 1160 was first observed on 2010 October 30 UT in the standard
NICI Campaign observing modes (previously described in \citealt{pztel} and 
\citealt{cd35}).  In the combined angular and spectral
difference imaging (ASDI) mode, where we are more sensitive to methane-bearing
planets, images are taken simultaneously in the 4\% bandpass 
methane on (1.652 $\mu$m) and off (1.578 $\mu$m) filters 
using NICI's two cameras.  In the ADI mode, 
where all of the light is sent to one 
camera, we are more sensitive to faint 
companions at separations $\gtrsim$2$''$.  We obtained 45 1-minute 
exposure images in the ASDI mode and 20 1-minute exposure images in the
ADI mode. In the ASDI mode, the primary is not saturated as it is observed 
through a partially transmissive coronagraphic mask at NICI's first focal 
plane.

We easily detected two companions at separations of 0.78$\pm$0.03'' and 
5.15$\pm$0.03''. After using archival VLT data 
to confirm that both companions are common proper motion 
(Sections \ref{archive_sec} and \ref{astrometry_sec}), 
we observed HD 1160 again on UT November 22, 2010 in the $J$, $H$ and $K_S$ 
bands (filters are in the MKO system) 
in order to measure colors for the companions.  Ten images 
were taken simultaneously in the $H$ and $K_S$ bands with exposure times of
12.16 seconds and 5 coadds per image. 
In the $J$-band, we obtained twenty images in 
single channel mode with the same exposure times and coadds per image.
Both companions were well detected even in individual images, but we chose 
only the best 7--10 images where the primary was centered most accurately on 
the mask.  We obtained an additional NICI dataset on 
UT October 21, 2011 (in the narrowband methane filters), which provided 
further astrometric confirmation.

\subsection{Keck-II NIRC2: $L^{\prime}$ and $M_S$ band imaging}

We observed HD 1160 in the $L^{\prime}$ (3.426--4.126 $\mu$m) and $M_S$ 
(4.549--4.790 $\mu$m) bands with the Keck II Telescope adaptive optics 
system and facility near-IR camera NIRC2 on UT 
November 27, 2010 without any coronagraphic mask.  We used the narrow 
camera which has 9.961 mas pixels \citep{ghez_nirc2}.  
In the direct imaging mode (non-ADI), we obtained 37 images at 
$L^{\prime}$ band, each with an exposure time of 0.15 seconds and 200 
coadds. In the $M_S$ band, we obtained 29
images with an exposure times of 0.2 seconds and 50 coadds. For each band, we 
used 5 dither positions with separations of 1$''$.  The median combination of 
the unshifted images was used to make 
sky images for subtraction. Both companions were clearly
detected and the primary was unsaturated in both bands.  In addition to the 
plate scale, we also use the Keck 
NIRC2 field rotation derived by \citet{ghez_nirc2} to 
determine the astrometry of the two companions.

\subsection{VLT NACO and ISAAC: $K_S$, $L$, and $L^{\prime}$ imaging}\label{archive_sec}

We examined archival data of HD 1160 taken at the VLT between 2002 and 
2012 in order to assess whether HD~1160~B and C are 
common proper motion companions to HD~1160~A.  These datasets include 
$K_S$ and $L^{\prime}$ (3.49--4.11 $\mu$m) 
adaptive optics images taken with the NACO system, and 
ISAAC $K_S$ and $L$ (3.49--4.07 $\mu$m) 
images, originally obtained for photometric 
calibration of observations of other science targets.  NACO images 
taken in the $K_S$-band have 27.06 mas pixels, while $L^{\prime}$ images have 
27.12 mas pixels. ISAAC has a $K_S$ pixel size of 148 mas and $L$-band images 
have a pixel size of 71 mas.  Despite their relatively short integration 
times (of order one minute per epoch), 
HD~1160~C ($\sim$5'' separation) is cleanly detected in all the data.  
HD~1160~B ($\sim$0.8'' separation) is only detected in the adaptive optics 
datasets and is not resolved in the seeing-limited ISAAC data.  Position 
angles measured with the NACO 
instrument are corrected to account for a $-$0.6 degree field rotation 
measured by \citet{bergfors11}.

\begin{figure}
\epsscale{0.9}
\plotone{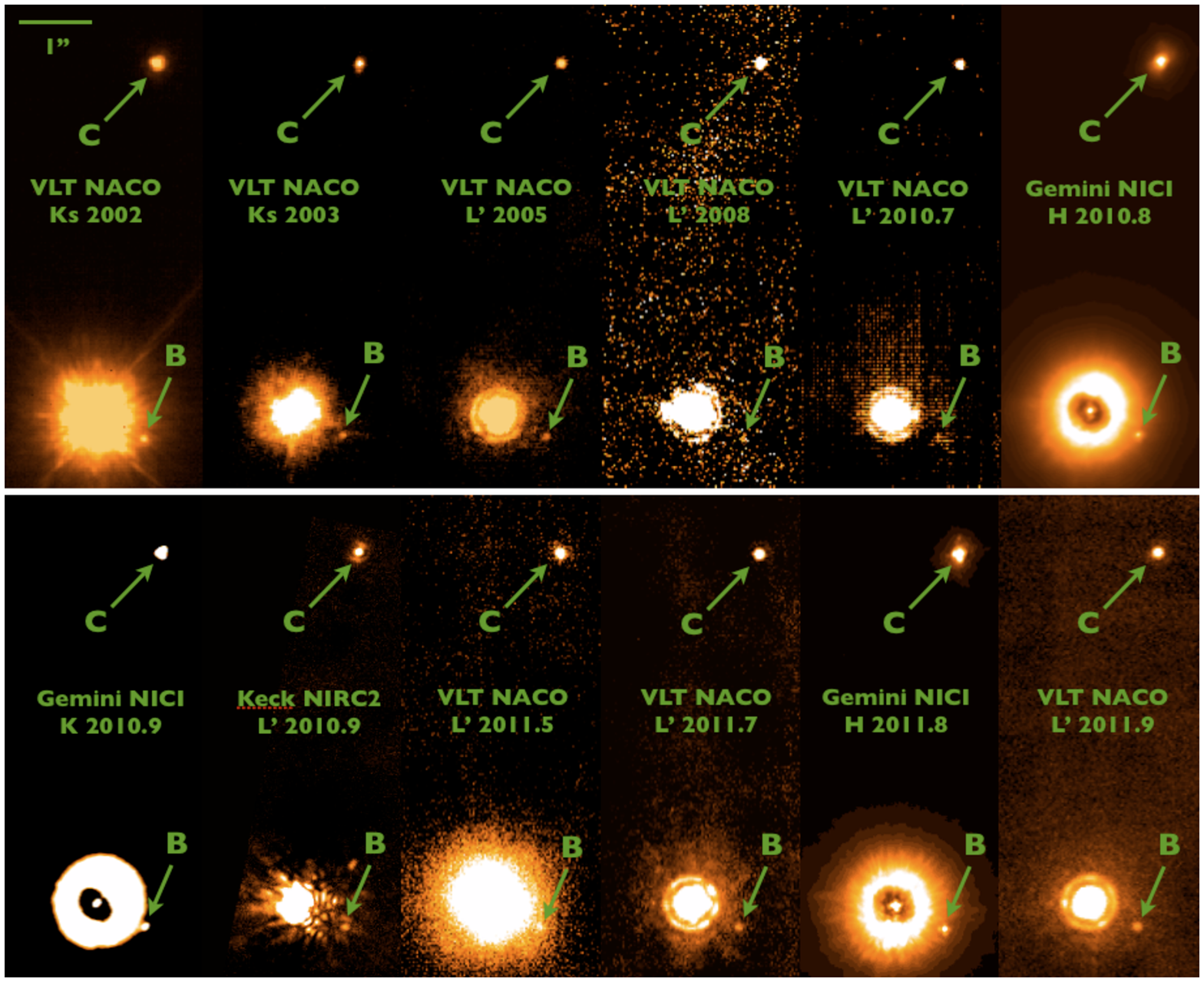}
\caption{Images of HD~1160~A, B, and C over twelve epochs 
from 2002 to late 2011.  The VLT NACO data were taken as 
calibration data for science programs targeting other objects, and in some 
cases HD~1160~B is detected at low S/N.  Each image was rotated 
to place North up and East to the left and has the same field of view, 
2.7''$\times$ 6.5''.
\label{images_abc}}
\end{figure}

\begin{figure}
\epsscale{0.9}
\plotone{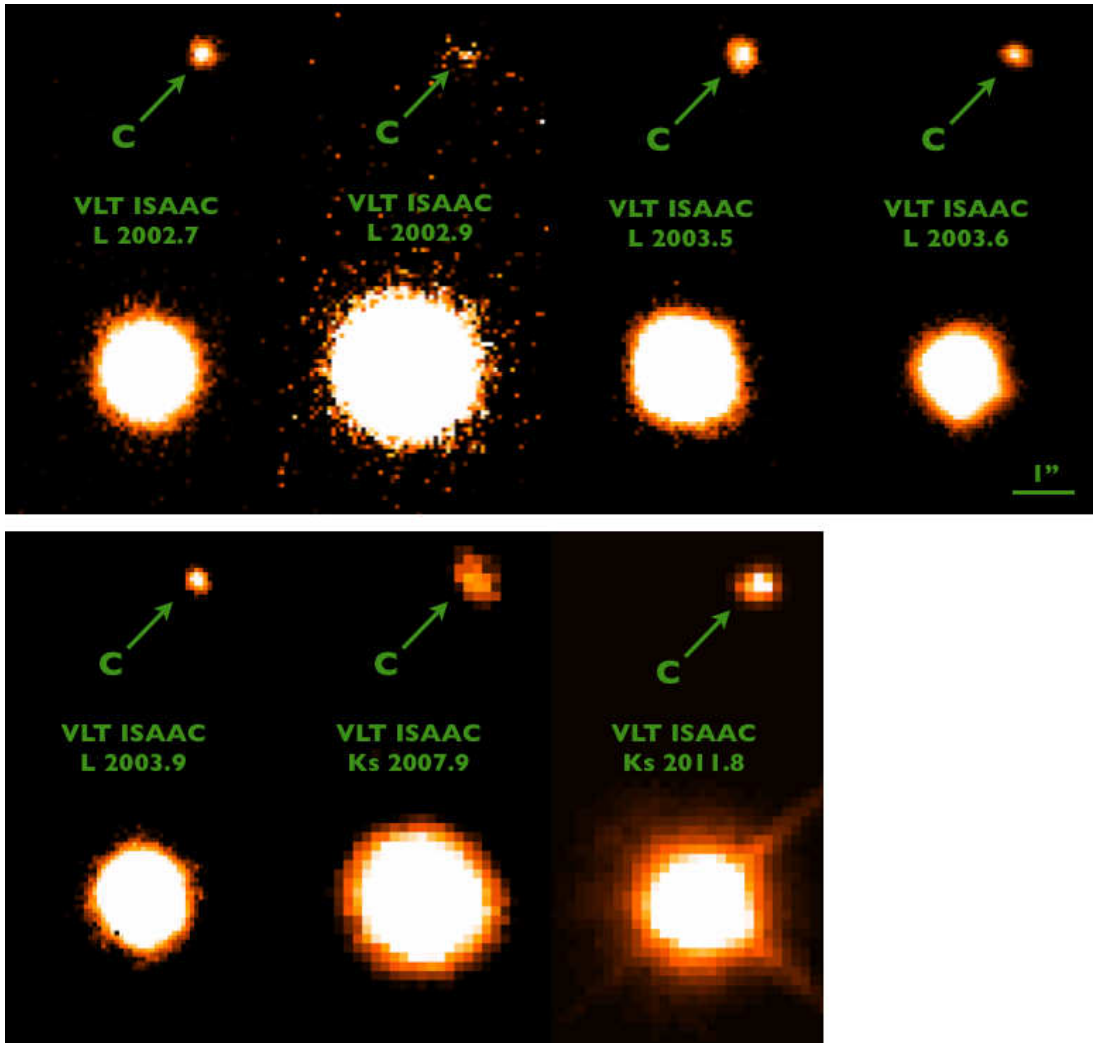}
\caption{Archival images of HD~1160~A and C from 2002 until 2011, taken with 
the VLT ISAAC instrument.  As with the VLT NACO data, these are 
calibration data and are sometimes at low S/N.  Additionally, since the 
seeing-limited resolution of ISAAC is lower than the AO-corrected images in 
Figure~\ref{images_abc}, HD~1160~B is not detected at any of these epochs.  
Each image was rotated to place North up and East to the left and 
has the same field of view, 2.7'' $\times$ 6.5''.
\label{images_ac}}
\end{figure}

Figure~\ref{images_abc} shows the 
detection of both companions at each VLT NACO, Gemini NICI, and Keck NIRC2 
epoch.  Figure~\ref{images_ac} shows the detections of only HD~1160~C at 
each of the seven ISAAC epochs.  Astrometry for all epochs is presented in 
Table~\ref{table2}.

\subsection{Magellan MIKE: Optical Spectroscopy}\label{mikesec}

We acquired a spectrum of HD~1160~A and C on UT 30 December 2010 
with the Magellan Inamori Kyocera 
Echelle (MIKE) spectrograph at the Magellan Clay telescope at Las Campanas 
Observatory in Chile. We used the 0.5$\arcsec$ 
slit which produces a spectral resolution of $\approx$35,000.  These 
data were reduced using the facility pipeline \citep{kelson03}.  We do 
not attempt to recover the spectrum of the much closer companion, HD~1160~B, 
with these observations.

Each stellar exposure is bias-subtracted and flat-fielded for 
pixel-to-pixel sensitivity variations. After optimal extraction, the 1-D 
spectra are wavelength calibrated with a thorium-argon arc. To correct for 
instrumental drifts, we used the telluric molecular oxygen A band (from 
7620--7660 \AA) which aligns the MIKE spectra to 40 m~s$^{-1}$, after 
which we corrected for the heliocentric velocity. The final spectra are of 
moderate S/N reaching $\approx$ 50 per pixel at 8700 \AA. We also observed 
GJ 908 (Spectral Type M1) as a radial velocity (RV) standard.

To measure the RV of HD~1160~C, 
we cross-correlated each of 9 orders
between 7000 and 9000~\AA\/ (excluding those with strong telluric absorption)
 with the spectrum of GJ 908
 using IRAF's\footnote[1]{IRAF (Image
  Reduction and Analysis Facility) is distributed by the National Optical
  Astronomy Observatories, which is operated by the Association of
  Universities for Research in Astronomy, Inc.~(AURA) under cooperative
  agreement with the National Science Foundation.} {\it fxcor} routine
(\citealt{fitz93}). We use GJ 908's $-$71.147~km~s$^{-1}$ RV published by the 
California and Carnegie Planet Search \citep{nide02}. The zero-point of the 
absolute RV is uncertain at the 0.4 km~s$^{-1}$ level. We measured the 
RV of HD~1160~C from the gaussian peak fitted to the
cross-correlation function (CCF) of each order and adopt the average RV of all
orders with the standard deviation of the individual measurements.  We 
measure the radial velocity of HD~1160~C to be 12.6 $\pm$ 0.3 km/s.  We also 
observed HD~1160~A itself with the MIKE spectrograph.  Given the 
few spectral lines available we were unable to get as precise an RV 
measurement as we did for the M dwarf HD~1160~C.  However, measuring the 
RV of several lines in HD~1160~A and taking the average, we found that 
the RV of HD~1160~A is 14 $\pm$ 2 km/s, consistent with that of HD~1160~C.

\subsection{IRTF SpeX: Near-IR Spectroscopy}

We obtained moderate-resolution ($R\approx$ 2000) near-IR spectra of
HD~1160~A and C on 29~June~2011~UT from NASA's Infrared Telescope Facility
(IRTF), located on Mauna Kea, Hawaii. Conditions were photometric with
good seeing conditions ($\approx$0.5--0.6\arcsec), and the C companion was
well-resolved from the primary star.  We did not attempt to recover 
the spectrum of the much closer companion, HD~1160~B, with these 
observations.  

We used the facility near-IR spectrograph
SpeX \citep{spex98} with the 0.3\arcsec\ wide slit in
cross-dispersed mode, obtaining spectra from 0.8--2.5~\micron.  
The slit was oriented at a position angle of 50$^{\circ}$ on the sky,
compared to the parallactic angle of $\approx$80$^{\circ}$  at the time, in
order to minimize the light from the primary that entered the slit. In
the pairwise-subtracted images, in the $H$ and $K$ bands, 
no evidence was seen of contaminating
light from the primary compared to the very well-detected companion.  Some 
contamination was seen in the $J$ band.  
Since the target was nearly overhead at the time of the observation 
(airmass ranging from 1.07 to 1.10), the
effect of atmospheric dispersion on the broadband spectrum is 
minimal.  The slit was not oriented perpendicular to the primary-companion 
axes,
since this would cause the companion's spatial position in the slit to
 coincide with a local maximum in the light from the primary's halo.
                                   
HD~1160~C was nodded along the slit in an ABBA pattern for a total 
on-source integration of 16~min, with individual exposure times
of 120~sec. We observed the primary star HD~1160 itself for telluric
calibration, since it has a spectral type of A0V and was well-resolved
from the companion. All spectra were reduced using version~3.4 of the
SpeXtool software package
\citep{vacca03,cushing04}. The median S/N per pixel in
the final reduced spectrum is 25.

\section{Results}

\subsection{Astrometric Confirmation of HD~1160~B and C}\label{astrometry_sec}

\begin{figure}
\epsscale{0.7}
\plotone{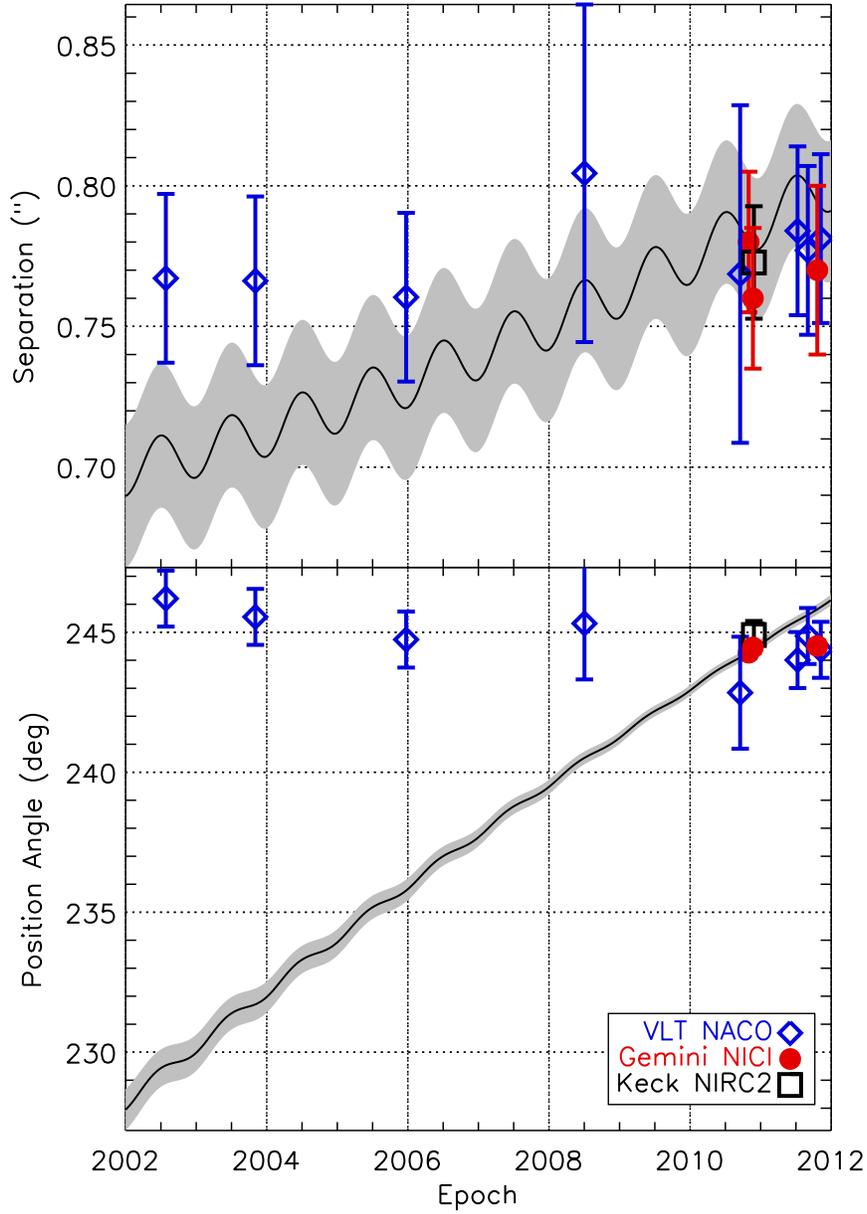}
\caption{Astrometry of HD~1160~B from 2002 to 2011 (points with error bars) 
along with the expected motion of a background object (black line).  A 
background object should fall within the gray shaded track, given the 
known proper motion and parallax of HD~1160~A.  The thickness of the gray 
track indicates the positional uncertainties arising 
from the proper motion and parallax uncertainties for HD~1160~A, 
as well as the astrometric uncertainties from the 
NICI 2010.8301 position to which the track is tied.  HD~1160~B is 
clearly a common proper motion companion.
\label{astrometryb}}
\end{figure}

\begin{figure}
\epsscale{0.7}
\plotone{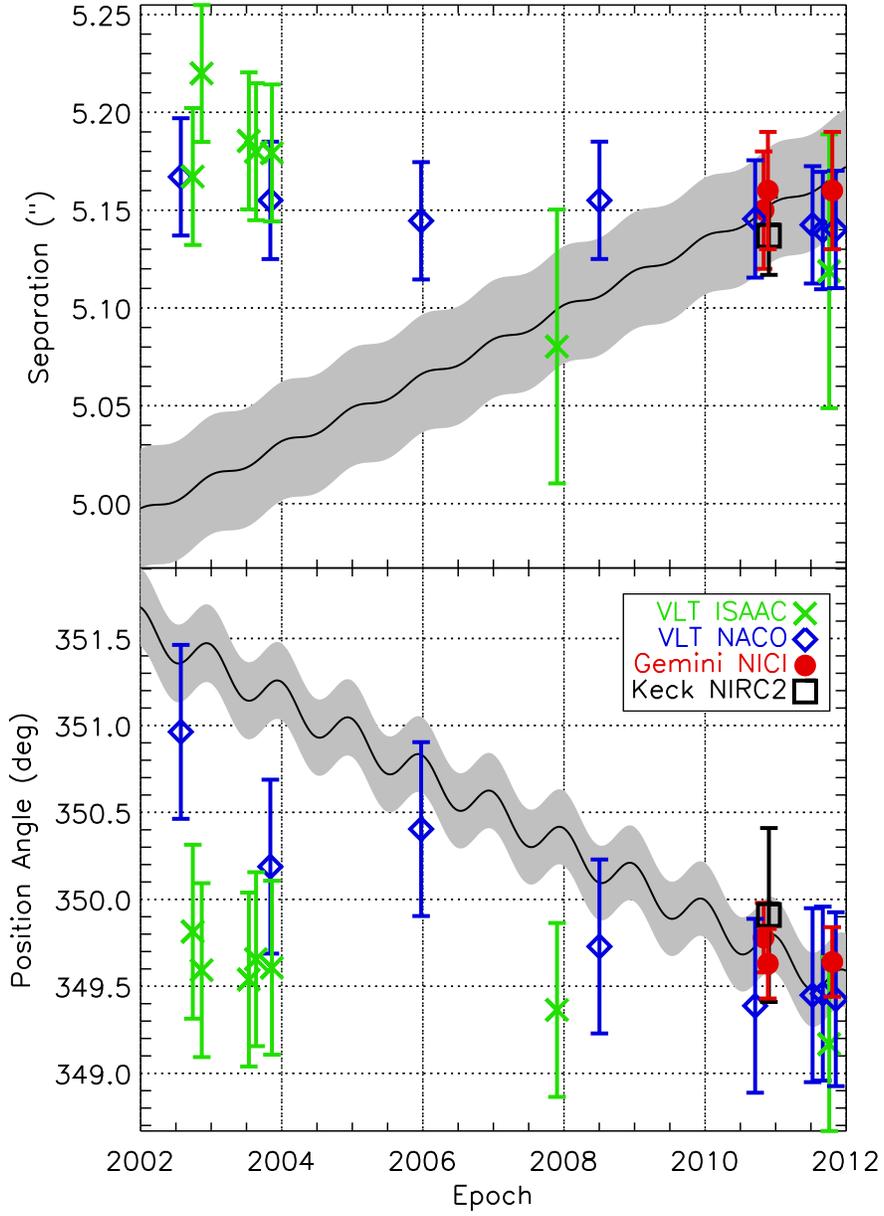}
\caption{Astrometry of HD~1160~C from 2002 to 2011.  While a less significant 
detection of common proper motion than with HD~1160~B, the data are still much 
more consistent with HD~1160~C being a co-moving companion rather 
than a distant 
background object.  \label{astrometryc}}
\end{figure}

Our full set of astrometry from VLT, Gemini-South, and Keck is 
reported in Table~\ref{table2}.  Figures \ref{astrometryb} 
and \ref{astrometryc} show the relative positions of HD~1160~B and C 
with respect to A over nearly a decade (2002--2011), along with the 
expected relative motion of a distant background object, given the 
known proper motion and parallax of HD~1160~A.  Errors on this astrometric 
track are computed through a Monte Carlo approach, using errors from 
the reference (NICI, epoch 2010.8301) 
position, and errors in the proper motion 
and parallax of HD~1160~A from Hipparcos \citep{newhip}.

\begin{figure}
\epsscale{0.85}
\plotone{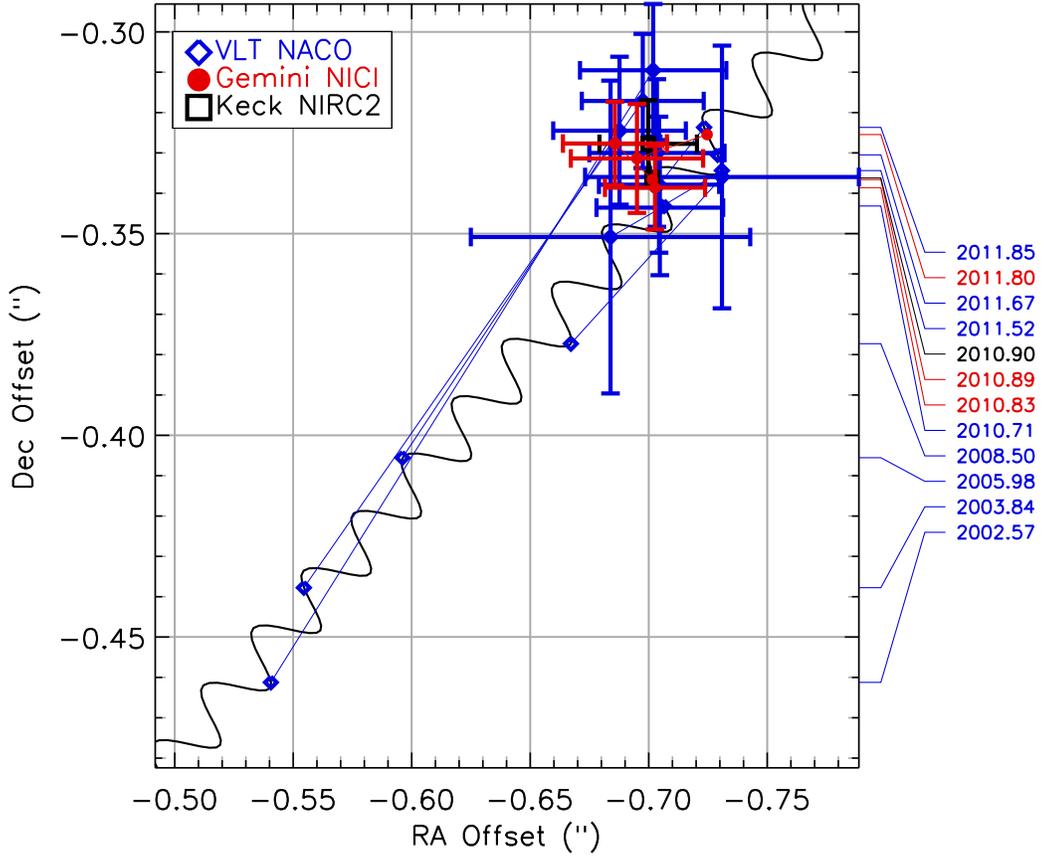}
\caption{Astrometry for HD~1160~B with respect to HD~1160~A as seen on the 
sky.  Each data 
point is connected by a thin line to the expected location of a 
distant background object at the observational epoch.  
The labels on the right give the epoch of the 
astrometric measurements, with the vertical location of the epoch label set 
by the Declination expected for a distant background object at the epoch.  
Within the error bars, all the datapoints are consistent with the 
position measured at the NICI 2010.8301 
detection and are significantly displaced 
from the expected positions for a distant background object.  The 
reduced $\chi^2$ value for HD~1160~B being a background object is 26.2 
(dof=22, P$\approx$0\%), compared to $\chi_{\nu}^2$ = 0.38 (dof=22, P=99.6\%) 
for a common proper motion companion.
\label{astrometry_paramb}}
\end{figure}

\begin{figure}
\epsscale{0.85}
\plotone{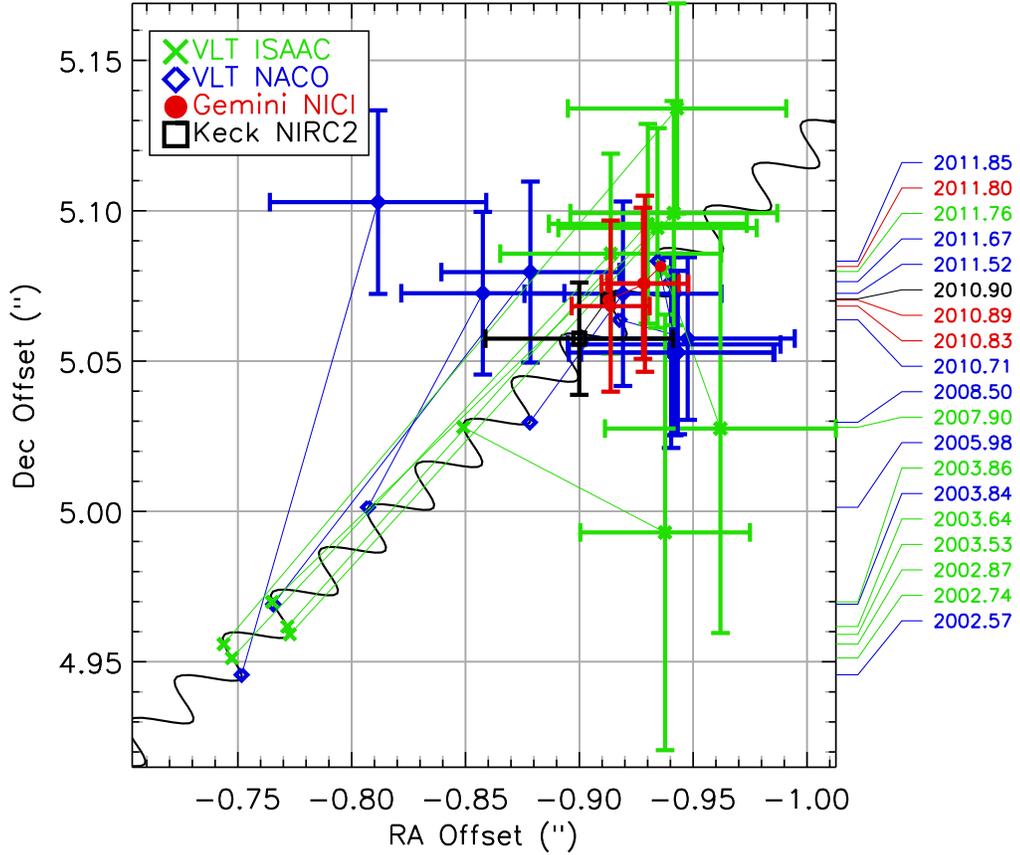}
\caption{On-sky astrometric plot for HD~1160~C with respect to HD~1160~A.  
The fit to the motion of a common proper motion companion is much more 
likely than the fit to the motion of a background object: the significance 
of the fit is 
$\chi^2_\nu$ = 4.95 (dof=36 P$\approx 0$\%) for a background 
object and $\chi^2_\nu$ = 0.62 (dof = 36, P=96.5\%) for a common proper motion 
companion.  In addition, the expected direction of the motion of a background 
object is orthogonal to the spread of the datapoints, and we have 
measured the radial velocities of HD~1160~A and C to be consistent within 
measurement uncertainty.
\label{astrometry_paramc}}
\end{figure}

Figures \ref{astrometry_paramb} and \ref{astrometry_paramc} show the 
astrometry of the companions on the sky.  
The astrometric measurements cluster on 
top of each other, as expected for common proper motion companions (CPM), 
and do not follow the background track.

To robustly determine whether 
each companion is more consistent with background or common proper motion, 
we compute the $\chi^2$ statistic using the equations

\begin{equation} \chi^2_{bg} = \sum_{i} \left ( \frac{(\rho_{obs,i} - 
\rho_{bg,i})^2}{\sigma_{\rho,i}^2} + \frac{(PA_{obs,i} - 
PA_{bg,i})^2}{\sigma_{PA,i}^2} \right )
\end{equation}

\noindent and 

\begin{equation} \chi^2_{CPM} = \sum_{i} \left ( \frac{(\rho_{obs,i} - 
\rho_{0})^2}{\sigma_{\rho,i}^2} + \frac{(PA_{obs,i} - 
PA_{0})^2}{\sigma_{PA,i}^2} \right )
\end{equation}

\noindent where $\rho_{obs,i}$ and $PA_{obs,i}$ are the observed separation 
and position angle at each epoch~(i), $\rho_{bg,i}$ and $PA_{bg,i}$ are the 
expected separation and position angle at each epoch from the background (bg) 
track, $\rho_0$ and $PA_0$ are constant separations and position angles 
(corresponding to common proper motion), and $\sigma_{\rho,i}$ and 
$\sigma_{PA,i}$ are the measurement uncertainties in the observed separation and 
position angle.

HD~1160~B is easily confirmed as a co-moving companion, differing from 
the expected motion of a background object by 
15 degrees in position angle.  The fit to the 
background track across all epochs has reduced chi square 
$\chi^2_\nu$=26.2 (with 22 degrees 
of freedom).  These data are not consistent with the background hypothesis, 
which is ruled out at the P$\approx 0$\% level.  Meanwhile, the fit to 
constant separation and position angle (i.e. common proper motion) produces 
$\chi^2_\nu$=0.38 (dof=22, P=99.6\%).  The deviation of 
HD~1160~C's motion from the background track is less than that for B.  
Nevertheless the offsets are significantly more consistent with 
common proper motion ($\chi^2_\nu$=0.62, dof=36, P=96.5\%) 
than background ($\chi^2_\nu$=4.95, dof=36, P$\approx 0$\%).  Though 
examining Figure~\ref{astrometry_paramc} shows that while there is a 
spread to the 19 astrometric epochs and they do not clearly overlap within 
error bars, the direction of the spread is orthogonal to the expected 
motion of a background object (and probably due to orbital motion, as we 
discuss below.  In addition, we note that the radial velocity 
measurements of HD~1160~A and HD~1160~C are consistent within measurement 
uncertainties (see Section~\ref{mikesec}), as we would expect from 
physically associated companions.

Figure~\ref{astrometryc} suggests a small change in position angle over 
time for HD~1160~C, raising the possibility that we are observing 
orbital motion.  
Fitting a straight line with non-zero slope to the astrometry for HD~1160~C 
produces $\chi^2_\nu$=0.42 (dof=34, P=99.8\%), similar to the common proper 
motion fit, 
indicating that with our astrometric errors, we are unable to detect a 
deviation from non-zero relative motion.  Assuming a mass for an A0 star of 
2.2 $M_{\sun}$ \citep{siess00}, 
a mass for HD~1160~C of 0.2 $M_{\sun}$ (see Section 
\ref{masssec}), and a face-on circular orbit with semi-major axis of the 
projected separation of 531 AU, we would expect orbital 
motion of $\sim$0.05 $^\circ$/year, while the best-fit line to the 
PA motion gives $-$0.04 $\pm$ 0.03 $^\circ$/year.  So while we do not 
significantly detect orbital motion, the magnitude of such motion would be 
consistent with that expected for HD~1160~C.

\subsection{Spectral Typing of HD~1160~C}

Figure \ref{kml_colmag} shows the $H$-$K$ and 
$K$-$L^{\prime}$ colors for HD~1160~B and 
C compared to field objects of M and L spectral types.  The 
locations of the HD 1160 companions in this color-color diagram 
suggest spectral types of mid-M for HD~1160~C, and  late M/early L for 
HD~1160~B.

Figure~\ref{hmk_colmag} shows HD~1160~B and C on an MKO NIR 
color-color diagram that includes field stars, field L dwarfs, selected 
individual low-mass objects, and members of the Taurus star-forming 
region.  We include in this plot photometry of field stars \citep{irtf2} 
and field L-dwarfs \citep{leggett10}, as well as the low mass objects 
AB~Pic~b \citep{abpic}, G~196-3~B \citep{rebolo98}, PZ~Tel~B \citep{pztel}, 
2M~1207~A \citep{mohanty07}, and TWA~5B \citep{lowrance99}.  In addition, we 
plot $JHK$ photometry for a list of spectroscopically confirmed Taurus members 
that do not possess circumstellar disks \citep{luhman10} or known binary 
companions \citep{kraus11,kraus12}.  The $JHK$ photometry (taken from 2MASS) 
has been dereddened using the extinction values from the compilation of 
\citet{kraus09} or the discovery survey, using the relative extinctions 
from \citet{reddening}, and converted to MKO using the conversion of 
\citet{leggett06}.

\citet{irtf2} note that this figure bifurcates for field dwarfs and 
giants with M spectral types; and indeed, while some giants are on the lower 
sequence, no field M dwarfs are found on the upper sequence.  HD~1160~B and 
C fall clearly on the upper branch with the M giants, and not the lower 
branch with the M dwarfs.  HD~1160~B and C cannot be giant stars themselves, 
as this would be inconsistent with their absolute magnitudes.  A small 
fraction of the Taurus objects lie on the upper sequence, but most lie on the 
lower branch.  That so few of these young ($\sim$2--5 Myr) objects fall 
on the upper branch appears to rule out the possibility that young objects 
(with intermediate surface gravity between giants and dwarfs) populate 
the upper branch and older objects the lower one.  Other plausible 
explanations for the 
five Taurus objects with giant-like colors include the presence of an 
undetected binary or disk, or an incorrect reddening determination.  Or, 
perhaps some small fraction of young stars lie on the upper branch, while most 
lie on the lower branch.  Alternatively, the discrepancy could also be 
accounted for by an error in our photometric reduction: shifting our NICI 
photometry of HD~1160~B and C by $\sim$2$\sigma$ would also place both 
objects on the lower track.  In any case, with our current data we are unable 
to explain why the $JHK$ photometry of HD~1160~B and C diverges from the 
photometry of dwarfs and the bulk of the single-star diskless Taurus objects.

To assign a spectral type for HD~1160~C, we plot our IRTF/SpeX 
spectrum against M-dwarf standards from 
\citet{irtf_mlt} in Figure~\ref{spectrac_jhk}.  We match flux levels for 
the entire $JHK$ spectrum using 
only the $H$ and $K$ portions of the spectra, since our $J$-band data is 
particularly noisy and likely contaminated by HD~1160~A.  
The best match to the shape of the continuum and 
spectral features comes from the M3 and M4 templates.  
As a result, we assign a spectral type of 
M3.5 $\pm$ 0.5 to HD~1160~C.

We have also compared the spectrum of HD~1160~C to an M3.5 III spectrum 
from \citet{irtf_mlt}, but find a poor fit to the IRTF/SpeX spectrum, for both 
the shape of the continuum and strength of the absorption features.  This is 
as we would expect: HD~1160~C, while having NIR colors similar to a giant, is 
not a giant itself.

Figure~\ref{spectrac_ew} compares the equivalent widths of atomic features 
for HD~1160~C with those for stars from 
\citet{irtf2} and members of the 
$\sim$10 Myr TW Hydra Association \citep{twa_spec}.  As in \citet{irtf2}, we 
follow the method of 
\citet{irtf_mlt} to compute equivalent widths and associated errors (see their 
Section 3.4), by defining continuum regions, fitting a polynomial, and 
directly computing the flux difference between measured spectrum and estimated 
continuum in the line regions.  A spectral 
type of M3.5V is consistent with HD~1160~C.  We cannot use these 
data to constrain the age, since at a spectral type of M3.5, there is 
not a significant offset between the field M dwarfs and young M stars in 
TWA in the strength of the absorption lines.

\subsection{Age of the HD 1160 System}

Determining the age of the HD 1160 system is challenging, since 
the most widely used 
age diagnostics (such as lithium absorption, calcium emission, H$\alpha$ 
emission, and rotation rate) 
are suitable only for stars with 
spectral types of F, G, and K \citep[e.g.][]{soderblom2010}.  In the 
case of HD~1160~ABC, with spectral types A, L, and 
M, none of the components are amenable to these techniques for 
determining ages, and so 
we turn to an analysis of the system's position on the HR diagram 
to estimate its age.  In fact, this is a 
more fundamental method of determining ages, since lithium and other 
indirect age indicators are calibrated by measurements from open clusters, 
whose ages are determined by comparison to isochrones on the HR diagram.

HD~1160~A was originally selected as a target for the Gemini 
NICI Planet-Finding Campaign due to its faint absolute magnitude for 
its spectral type.  \citet{jura98} and \citet{lowrance00} note 
that on the HR diagram early-type 
stars in young clusters tend to be lower (i.e. have smaller 
luminosities) than stars in older clusters.  In fact, 
\citet{su06} flag HD~1160~A as lying below the zero-age main sequence, 
implying an age of $<$5 Myr.  (HR 8799, the host of 4 directly imaged planets, 
was determined to be young by the same method [\citealp{hr8799b}].)  With 
our discovery of two co-moving companions 
to HD~1160~A, we can expand this HR diagram analysis to include all 
three components, with spectral types ranging from A0 to early L.

Figure \ref{ageplot1} shows HD~1160~A and C compared to stars at six 
sets of ages: Upper Sco \citep{kraus_usco} at 5 Myr; 
members of $\sim$10 Myr ($\beta$ Pic and TW Hya) and 
$\sim$30 Myr (Tuc/Hor, Carina, and 
Columba) moving groups \citep{torres08}; the 120 Myr Pleiades open cluster 
\citep{stauffer07}; the Hyades open cluster 
\citep{roeser11} at 600 Myr; and low-mass field objects from the 
Palomar/MSU Nearby-Star Spectroscopic Survey \citep{reid96}.  
All NIR photometry of reference objects were obtained with 2MASS filters, 
and so we use the conversion factors of 
\citet{leggett06} to convert the HD~1160~ABC MKO magnitudes into 
2MASS magnitudes for Figure~\ref{ageplot1} (we do the same for 
Figure~\ref{ageplot2}).  As we currently
lack spectra for HD~1160~B and cannot determine its spectral type, 
we do not plot it in Figure~\ref{ageplot1}.  
From these plots, the HD 1160 system appears younger 
than the Hyades (given the low luminosity of HD~1160~A with respect to the 
Hyades A star sequence), but consistent with the younger 
associations.

In Figure \ref{ageplot2}, we show a NIR color-magnitude diagram for the 
same associations, along with all three members of the HD 1160 system.  Here, 
HD 1160 again appears younger than the Hyades and the field objects, and 
inconsistent with the 
Pleiades as well, with HD~1160~A below the A-star sequence, and HD~1160~B and 
C redder than M stars of similar absolute $J$ magnitudes.  All three 
components are consistent with similar objects in Upper Sco or the 10 and 
30 Myr 
moving groups (though the paucity of known low-mass members to the two sets of 
moving groups makes a detailed comparison at these ages difficult).  
As a result, we set an upper limit for the age of the HD 1160 
system of 100 Myr, and adopt an age range of 
50$^{+50}_{-40}$ Myr.

\subsection{Metallicity}

An alternate explanation for the underluminosity of HD~1160~A is 
that the system has a sub-solar metallicity, as we illustrate in 
Figure \ref{metplot1}.  While HD~1160~A lies below the \citet{siess00} 
tracks at [Fe/H] = 0.0, its position on the color-magnitude 
diagram is consistent for ages up to 300 Myr at [Fe/H] = $-$0.3, half the 
metal abundance of the Sun.  Similarly, \citet{merin04} found that the 
low position of HD 141569 (a pre-main sequence B9.5V with a 
circumstellar disk) on the HR diagram as well as its 
medium-resolution optical spectrum could be explained by a metallicity of 
[Fe/H] = $-$0.5.

We note, however, that Figure \ref{ageplot2} provides two indications of 
youth for HD 1160: the underluminosity of A and the redness of B and C in 
$J-K_S$ color.  In Figure \ref{metplot2} we show a NIR color-magnitude 
diagram with HD~1160~B and C and field M dwarfs of known metallicity 
\citep{leggett00,rojas10}.  M dwarfs with subsolar metallicity have bluer 
$J-K_S$ colors than more metal-rich M dwarfs.  
So if the HD 1160 system were to be 
old ($>$100 Myr) but metal-poor, we would expect bluer $J-K_S$ colors 
for HD~1160~B and C than we observe.  The underluminosity of A stars on 
color-magnitude diagrams may indicate either young age or low 
metallicity, but the only available explanation for red $J-K_S$ colors in 
late-type stars is youth.  Indeed, \citet{johnson_koi254} derive a 
relationship between $J-K_S$ color and metallicity of M dwarfs, with redder 
$J-K_S$ color indicating more metal-rich atmospheres.

We also note that regardless of whether 
HD~1160~A has [Fe/H] = 0.0 or $-$0.3, the 
expected main sequence lifetime of an A0 star is $\sim$300 Myr 
\citep{siess00}, which provides an upper limit on the age of the 
system independent of the measured photometry of HD~1160~B and C.  
Additionally, since stars 
recently formed in the solar neighborhood tend to be of solar metallicity 
\citep{pedicelli09}, 
for an age less than $\sim$300 Myr, the HD 1160 system would have had to have 
formed further out in the galactic disk (i.e. in a lower metallicity 
environment) and migrated inward, which 
is inconsistent with its main sequence lifetime of $\sim$300 Myr and its 
low space motion (see Section \ref{smsec}).  Though 
a metallicity of [Fe/H] = $-$0.3 is at the lower limit of Cepheids at the 
solar radius, as measured by \citet{pedicelli09}, the more likely 
explanation is 
that HD~1160~A is solar metallicity, and its low position in the HR diagram 
is due to youth.

\subsection{Space Motion}\label{smsec}

By combining our measured RV of the HD 1160 system with the 
position and revised Hipparcos parallax and proper motion of HD~1160~A 
\citep{newhip}, we have computed the space motion of HD 1160.  We find 
the space motion to be $U$ = $-$7.6 $\pm$ 0.4, $V$ = $-$3.4 $\pm$ 0.5, 
$W$ = $-$15.7 
$\pm$ 0.4 km/s.  In Figure \ref{smplot}, we compare this space motion to those 
of young, nearby moving groups, as given by \citet{torres08}.  HD 1160 
is not co-moving with any known moving group, though its space motion is 
relatively low, with a total motion of 17.7 $\pm$ 0.5 km/s.  Such a 
low velocity is consistent with a young age for HD 1160.  We also note that 
this $UVW$ motion of HD~1160 places the system slightly outside (though near 
the edge) of the 1$\sigma$ $UVW$ ellipse of the ``young disk'' computed by 
\citet{eggen89} of $U$ = $-$15 $\pm$ 14, $V$ = $-$14 $\pm$ 9, 
$W$ = $-$6 $\pm$ 13 km/s. 

\subsection{Mass of HD~1160~B and C}\label{masssec}

Using the age and available photometry and spectroscopy 
of HD~1160~B and C, we proceed to assign masses to these companions using 
theoretical evolutionary models for stellar and sub-stellar objects.  
Table \ref{table4} shows mass estimates from the Lyon evolutionary 
models using the DUSTY atmospheric models \citep{chabrier00,dusty01} 
for HD~1160~B for the absolute magnitudes in each infrared band.  
Errors are calculated by interpolation across gridpoints of the 
DUSTY models, and using a Monte Carlo analysis assuming gaussian 
distributions for the magnitude uncertainties and a 
uniform distribution of ages from 10--100 Myr.  
Given the large age range, the implied mass range 
is quite large (24--90~M$_{Jup}$), 
spanning much of the range of brown dwarf masses.  
We note that these ranges only account for 
uncertainties in fluxes and the age, and do not account for any systematic 
uncertainty in the models themselves.  From the $J$-band (the peak of the NIR 
SED for late M and early L spectral types) flux alone, we 
derive a mass for HD 1160 B of 33$^{+12}_{-9}$ M$_{Jup}$.  Using the 
BC($J$) bolometric correction of \citet{liu10} (BC($J$) chosen since it is 
flat over the plausible range of spectral type for HD~1160~B, L0$\pm$2), we 
find a similar mass by computing the bolometric luminosity 
(L = 5.9$\times 10^{-4}$ $\pm$ 5$\times 10^{-5}$ L$_{\sun}$), and 
comparing to the DUSTY models: M = 37$^{+12}_{-10}$ M$_{Jup}$.

For HD~1160~C, we incorporate our measured spectral type of M3.5 $\pm$ 0.5 
when estimating the mass.  Using the dwarf sequence of \citet{luhman99}, we 
convert this spectral type to an effective temperature of 3270 $\pm$ 90 K, 
(or 3340 $\pm$ 70 K using the intermediate scale) and using a K-band 
bolometric correction from \citet{sptyperef2}, we 
compute a luminosity of HD~1160~C of 9.4$\times 10^{-3}$ $\pm$ 
8$\times 10^{-4}$ L$_{\sun}$.  
Coupled with the NextGen models of \citet{nextgen}, these 
values allow us to estimate a mass for HD~1160~C of 190$^{+65}_{-40}$ 
M$_{Jup}$.  Using the \citet{luhman99} intermediate sequence (more 
appropriate for younger stars), we find the mass of HD~1160~C to be 
230$^{+30}_{-45}$ M$_{Jup}$ (0.22$^{+0.03}_{-0.04}$ M$_{\odot}$). 

\subsection{Orbital Stability}

We estimate the likely semimajor axes of the two companions from their 
projected separations using the conversion factor of \citet{dupuy11}: 
assuming a uniform eccentricity distribution between 0 and unity and no 
discovery bias, their estimated ratio of semi-major axis ($a$) to 
projected separation ($\rho$) is 
$a / \rho = 1.10$ with 68.3\% confidence limits between 
0.75 and 2.02.  Given projected separations of 81 $\pm$ 5 AU and 
533 $\pm$ 25 AU for HD~1160~B and C, and using a Monte Carlo method to 
propagate errors (using a gaussian distribution for separation and 
two gaussians to account for the assymetric error 
bars on $a / \rho$), we estimate semi-major axes of 74 $\pm$ 
35 AU for HD~1160~B and 500 $\pm$ 200 for HD~1160~C.  

The HD~1160~ABC system appears to be orbitally stable, assuming that the
eccentricity of the A and C components is not larger than about 0.3.  Given 
the mass ratio between HD~1160~A and C of 0.1, and using the 
the detailed orbital stability calculations of \citet{holman99}, and assuming 
an orbital eccentricity of these two major components of 0.3, we
estimate that for the HD 1160 system the critical semimajor axis for 
stability of component B is about 0.2 times the semimajor axis of 
components A and C.  That is, component B can orbit no farther from A than 
about 100 AU. With an estimated semi-major axis of about 74 AU, component B 
therefore appears to be part of a stable
hierarchical system. Smaller orbital semimajor axes for HD
1160 B would also be stable, and so additional companions could plausibly 
exist interior to B.

Since we have not fit an orbital solution to our astrometry, 
the semi-major axes of both components may differ from our estimates.  
Indeed, if the semi-major axis is 109 AU for HD~1160~B and 300 AU for 
HD~1160~C (the 1$\sigma$ upper and lower limits of their 
respective semi-major axis estimates), then the system would 
not be orbitally stable.  Additionally, a high eccentricity for one or 
both of HD~1160~B and C would suggest instability.  Nevertheless, given 
our measurements, and the small probability of catching this system in an 
unstable state, the most likely explanation is that the system is stable.

\section{Conclusions}

We have described here the discovery and analysis of two low-mass companions 
to the young A star HD~1160~A, found during the Gemini NICI 
Planet-Finding Campaign.  With estimated companion masses 
of 33$^{+12}_{-9}$ and 230$^{+30}_{-45}$ M$_{Jup}$ and a primary mass of 
$\sim$2.2 M$_{\sun}$, this system has very small 
mass ratios between components, 0.10 between A and C, 
and 0.014 between A and B.  Around a star slightly less massive than the Sun, 
this A-to-B mass ratio would represent a planetary-mass companion.  

Hierarchical triple systems like HD 1160 occur with a frequency of about
8\% around solar-type stars \citep{raghavan10}, though seldom
with such extreme mass ratios as the A to B components of HD~1160.  Mass ratios
less than about 0.1 are rare (see Figure 16 of \citealt{raghavan10}) 
though not unprecedented, e.g., $\zeta$ Vir B, an M-dwarf companion to 
an A star, with mass ratio 0.08 \citep{zetavir} resembles HD~1160~C, while 
the brown dwarf HR 7329 B, with mass ratio of $\sim$0.01 relative to its 
A-star primary \citep{lowrance00}, is more akin to HD~1160~B. Along with binary 
star systems, multiple
star systems like HD~1160 are thought to form by the collapse and
fragmentation of dense molecular cloud cores (e.g., \citealt{boss09} and
references therein). The cloud core that collapsed to form HD 1160
presumably was relatively rapidly rotating, in order to result in the
wide orbits of all three components. Typical separations for binary
stars with mass ratios of about 0.1 (as for A and C) are in the range
of about 10 to 10$^4$ AU (see Figure 17 of \citealt{raghavan10}), so a
separation of A and C of about 531 AU is consistent with this range.

\begin{figure}
\epsscale{0.8}
\plotone{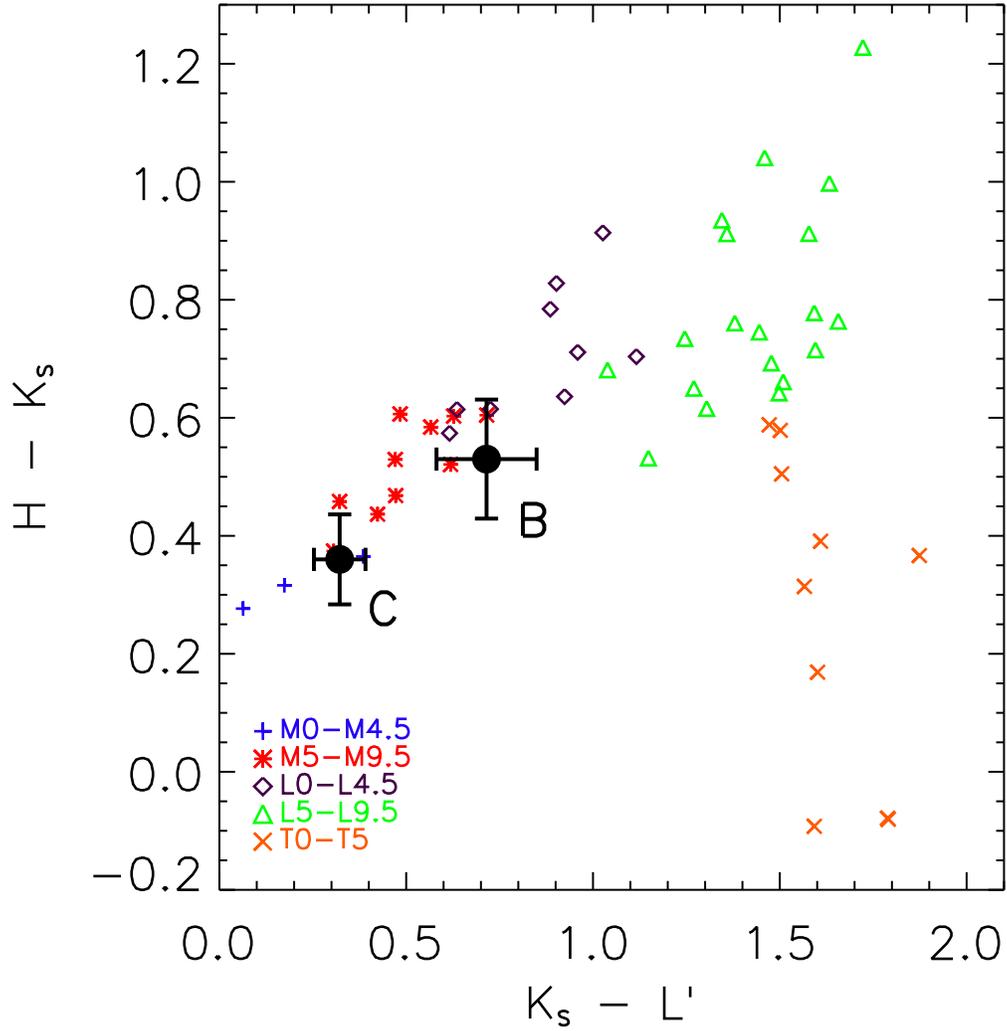}
\caption{Color-color diagram on the MKO system, 
suggesting spectral types of early-L 
and mid-M for HD~1160~B and C, 
respectively.  The low-mass field dwarfs plotted here are from the 
compilation of \citet{leggett10}.  The implied spectral type for HD~1160~C is 
a close match to our determination of M3.5 based on near-IR spectroscopy.
\label{kml_colmag}}
\end{figure}

\begin{figure}
\epsscale{0.8}
\plotone{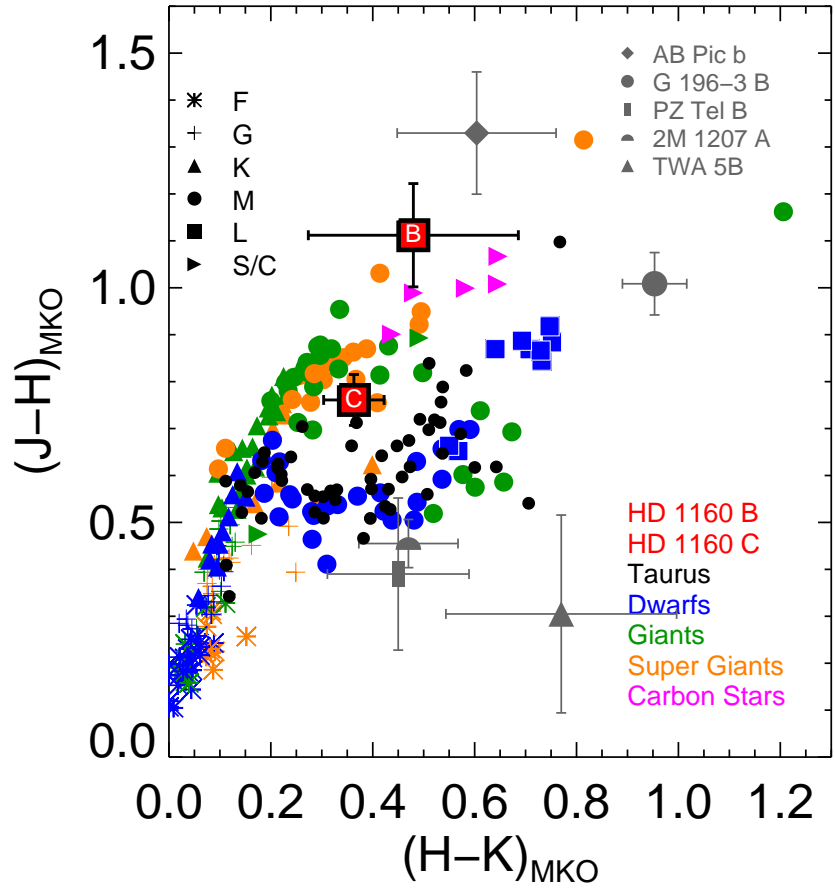}
\caption{Color-color diagram showing HD~1160~B and C, along with 
giants, dwarfs, single-star members of Taurus, and some individual 
low-mass objects.  While HD~1160~B and C are not giants themselves, they do 
exhibit giant-like $JHK$ colors.
\label{hmk_colmag}}
\end{figure}

\begin{figure}
\epsscale{0.95}
\plotone{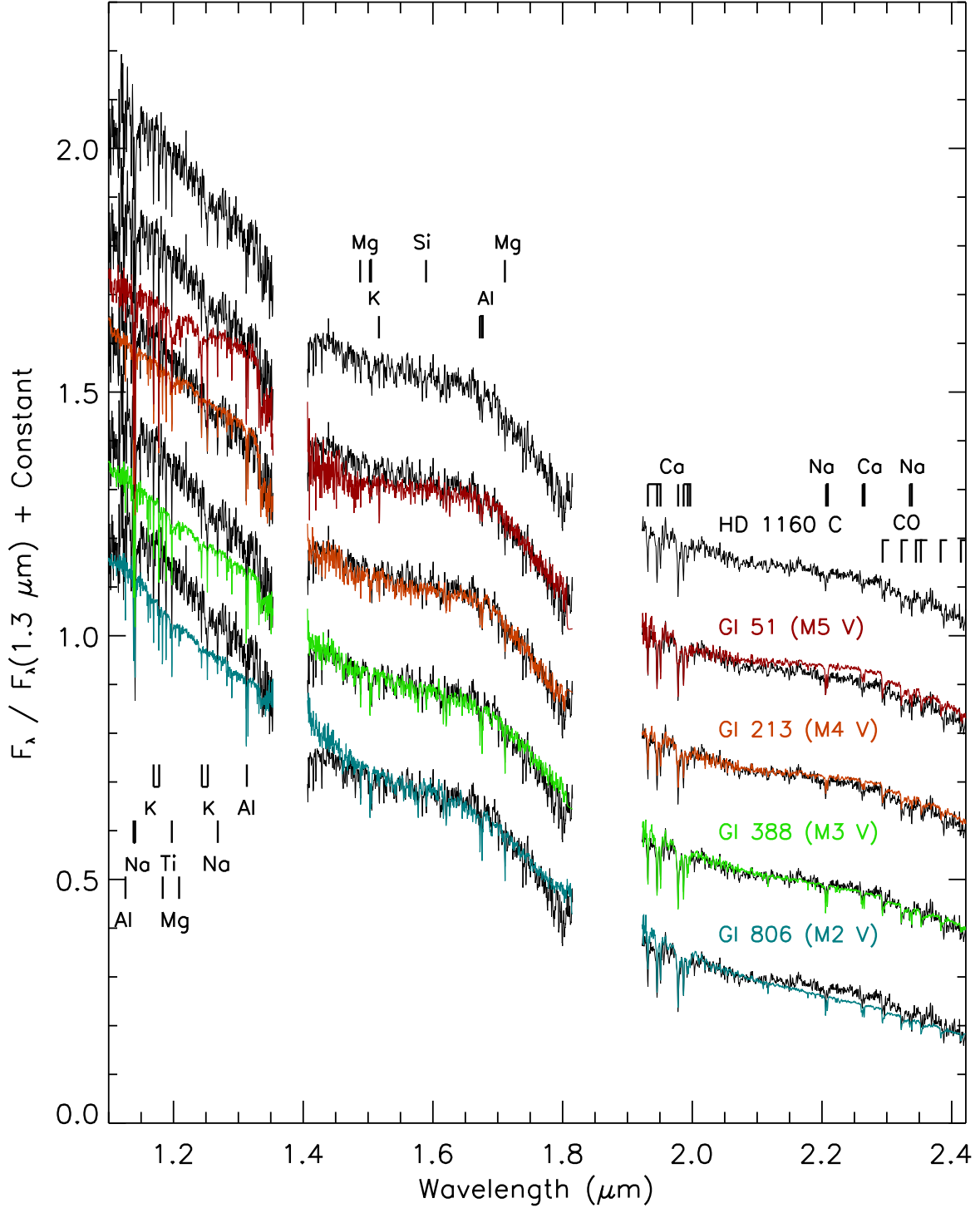}
\caption{IRTF/SpeX spectra of HD~1160~C compared to M dwarf spectral 
standards and line identifications from 
\citet{irtf_mlt}.  The best fits to the shape of the $H$ and $K$ continuum 
are spectral types M3 and M4, and therefore 
we assign a spectral type of M3.5.  
Note that the $J$-band portion of the spectrum suffers from 
contamination from HD~1160~A (A0), and so we only use the $H$ and $K$ 
portions in our fits.
\label{spectrac_jhk}}
\end{figure}

\begin{figure}
\epsscale{0.75}
\plotone{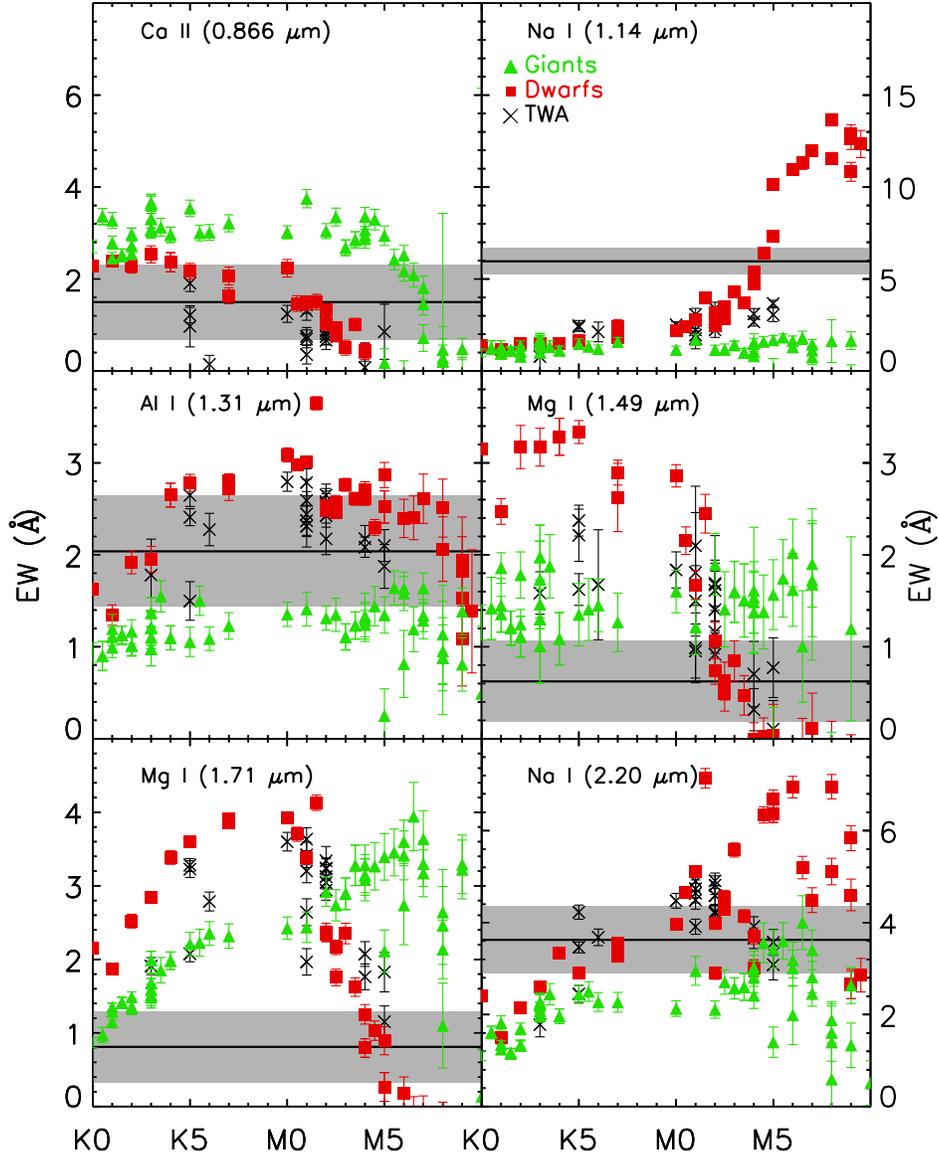}
\caption{Comparison of equivalent widths of atomic features in the HD~1160~C 
spectrum to those of dwarf and giant stars from \citet{irtf2} and members 
of the TW Hydra Association (TWA) from \citet{twa_spec}.  
Red squares denote dwarfs, green triangles giants, and black crosses 
TWA members.  For each 
feature, the black horizontal line denotes the equivalent width of 
HD~1160~C and the gray swath the corresponding measurement uncertainty.  
Taken together, the atomic absorption lines 
support our M3.5 spectral type for HD~1160~C.  However, 
these data cannot constrain the age of HD~1160~C, since the 
equivalent widths for M dwarfs and TWA members are similar.
\label{spectrac_ew}}
\end{figure}

\begin{figure}
\epsscale{1}
\plotone{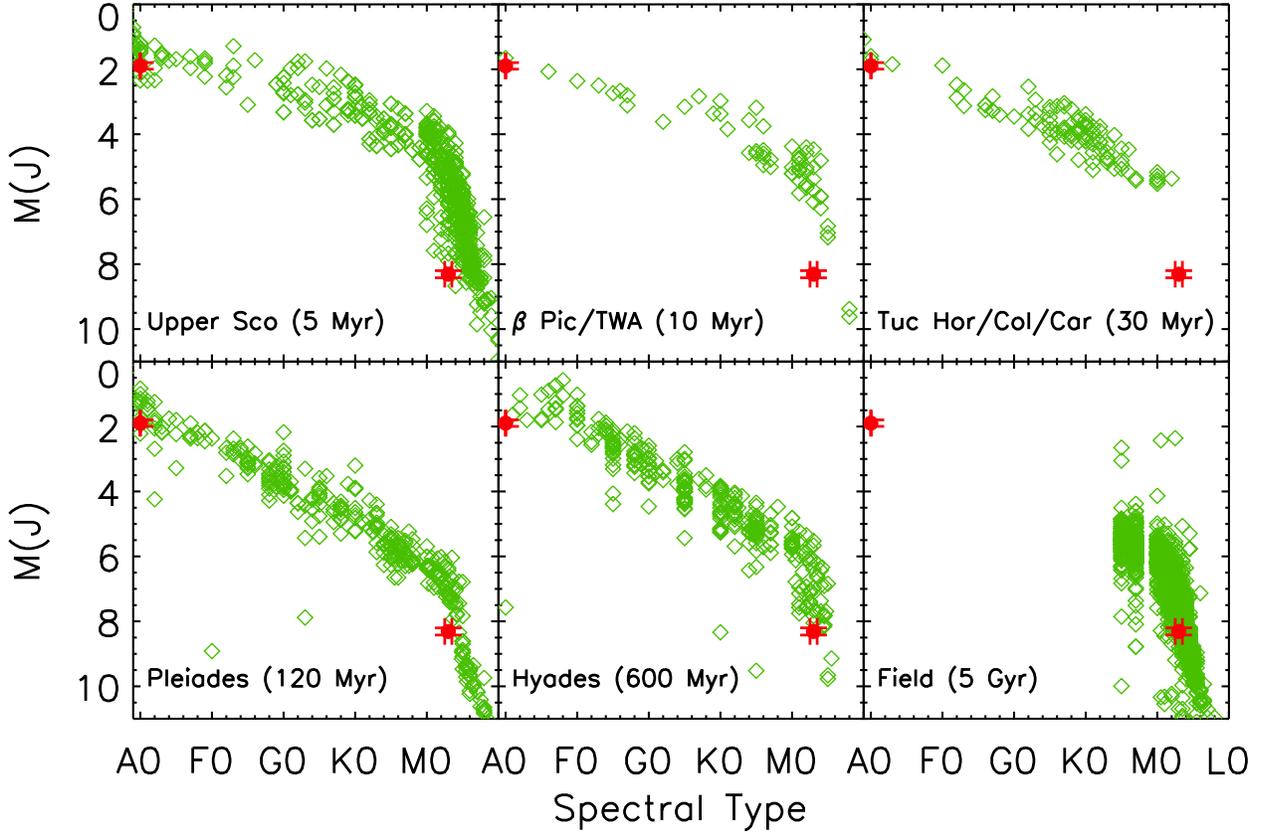}
\caption{HR diagram of stars in Upper Sco, 10 Myr moving groups ($\beta$ Pic 
and TW Hya), 30 Myr moving groups (Tuc/Hor, 
Carina, and Columba), the Pleiades, the Hyades, and field 
low-mass objects (magnitudes are in the 2MASS system).  The 
positions of HD~1160~A and C are marked with filled red circles.  We do not 
include HD~1160~B, as we do not have a spectral 
type.  HD~1160~A appears underluminous compared to other A stars, 
especially for older stars in the Hyades.
\label{ageplot1}}
\end{figure}

\begin{figure}
\epsscale{1}
\plotone{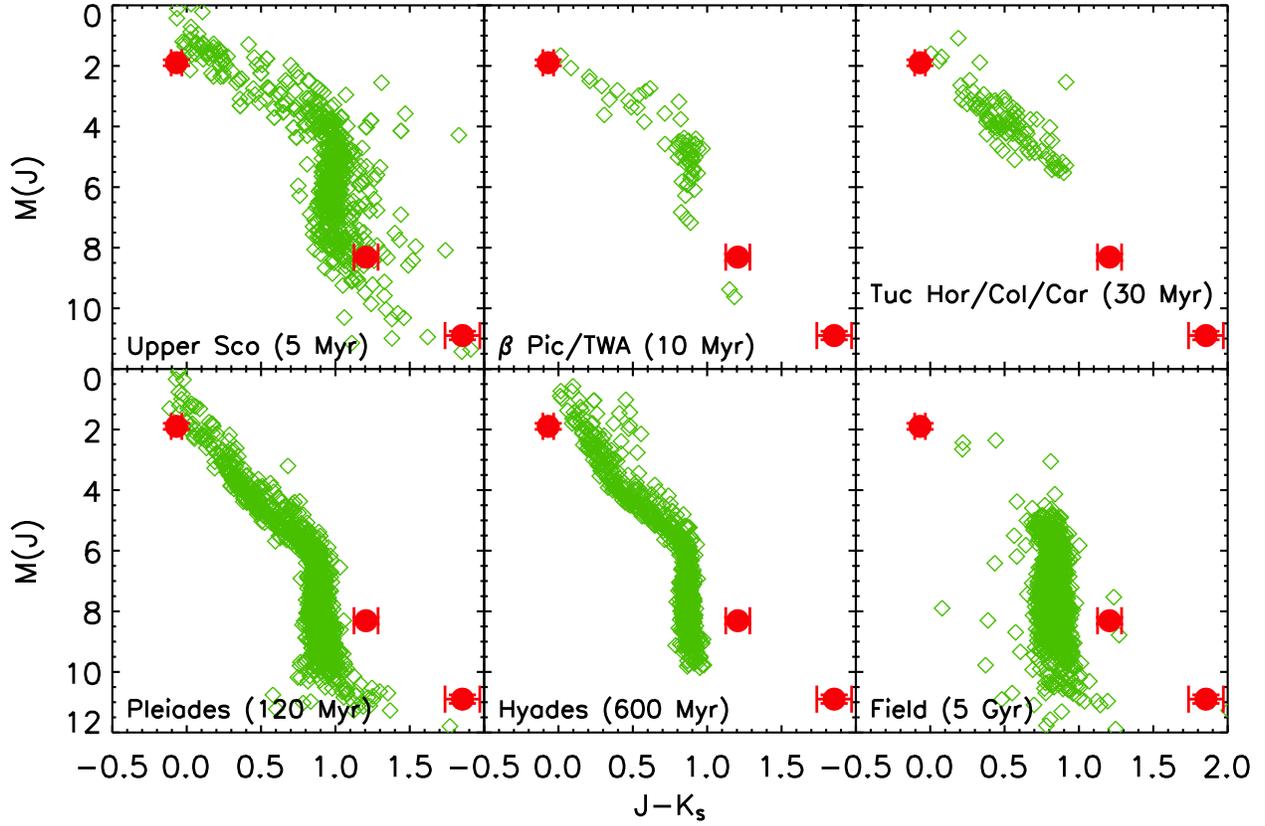}
\caption{Color-magnitude diagram (2MASS system) 
showing HD~1160~ABC with respect to stars 
in stellar associations of different ages.  HD~1160~A is too underluminous 
to fit either the Hyades or Pleiades stars, while HD~1160~B and C are too red 
to fit the low-mass end of the main sequence of these older open clusters or 
field objects.  The HD 1160 system does appear 
consistent with the stars in Upper Sco and the 10 and 30 Myr moving 
groups.  As a 
result, we assign an age to the system of 50$^{+50}_{-40}$ Myr.
\label{ageplot2}}
\end{figure}

\begin{figure}
\epsscale{1}
\plotone{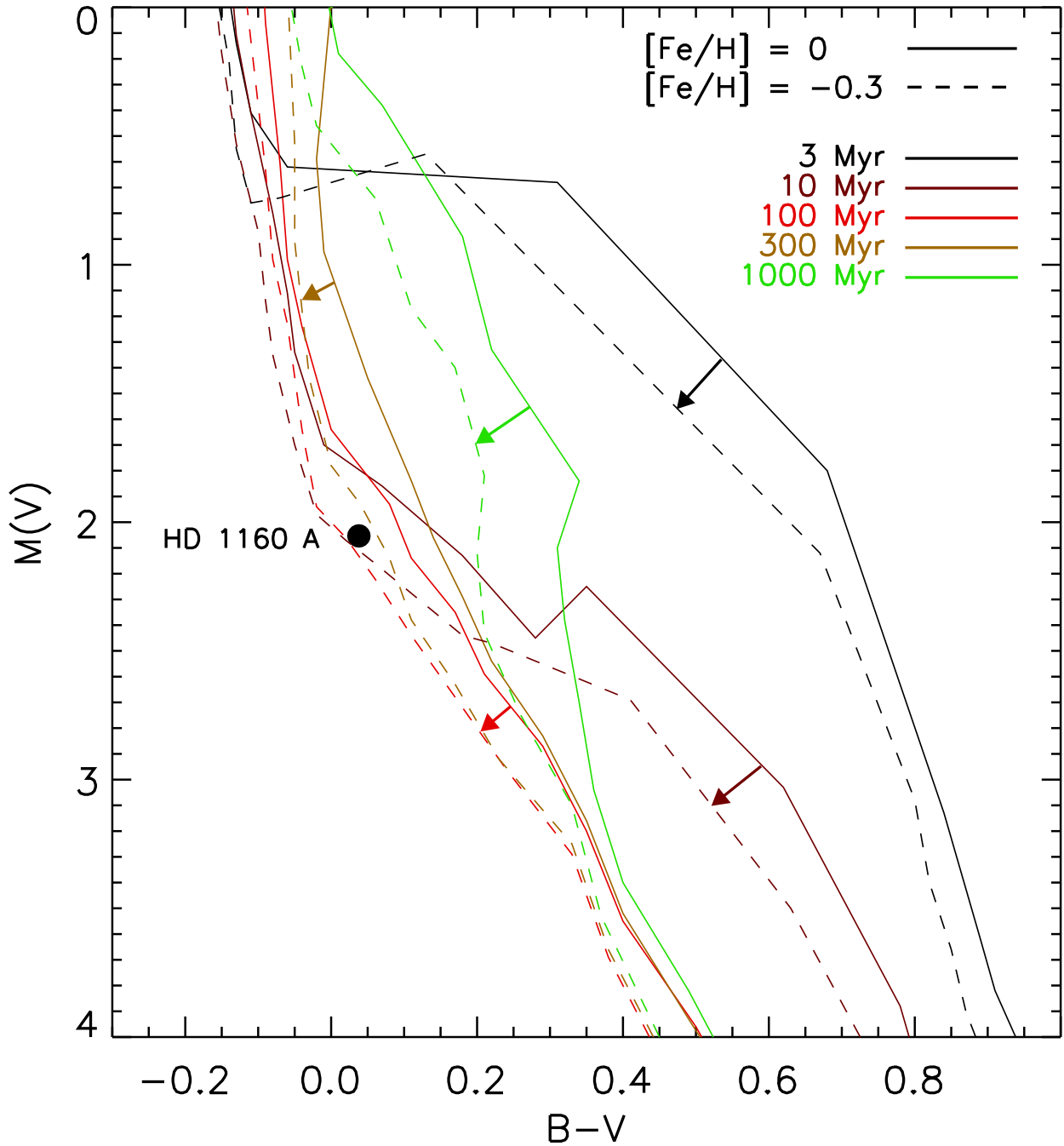}
\caption{Optical color-magnitude diagram showing the position of HD~1160~A 
(filled circle) against \citet{siess00} isochrones, for solar metallicity 
(solid lines) and half-solar abundance (dashed).  HD~1160~A is underluminous 
compared to tracks of all ages at [Fe/H] = 0.0, but is consistent with ages 
of 10--300 Myr for [Fe/H] = $-$0.3.  However, the colors of the B and C 
components do not agree with a low metallicity for the system (see 
Figure~\ref{metplot2})
\label{metplot1}}
\end{figure}

\begin{figure}
\epsscale{1}
\plotone{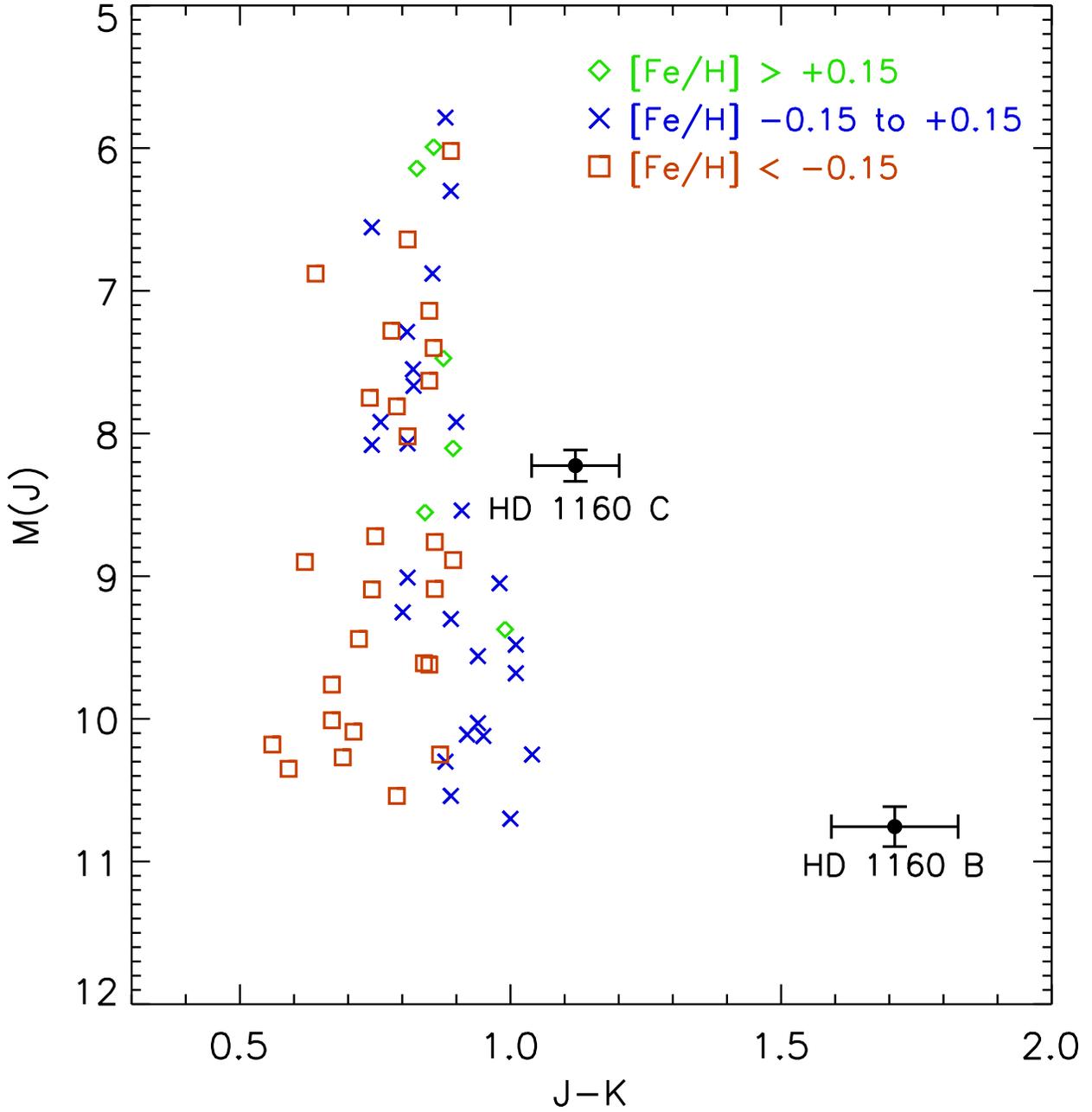}
\caption{HD~1160~B and C compared to 
M dwarfs of known metallicity, compiled by 
\citet{rojas10} and \citet{leggett00}.  Low-metallicity M dwarfs tend to be 
bluer in $J-K$ color than solar-metallicity dwarfs, especially at later 
spectral types.  So while the underluminosity of A stars such as HD~1160~A 
can be explained by either youth or sub-solar metallicity, the $J-K$ redness 
of HD~1160~B and C supports the idea that the system is young.  
\label{metplot2}}
\end{figure}

\begin{figure}
\epsscale{1}
\plotone{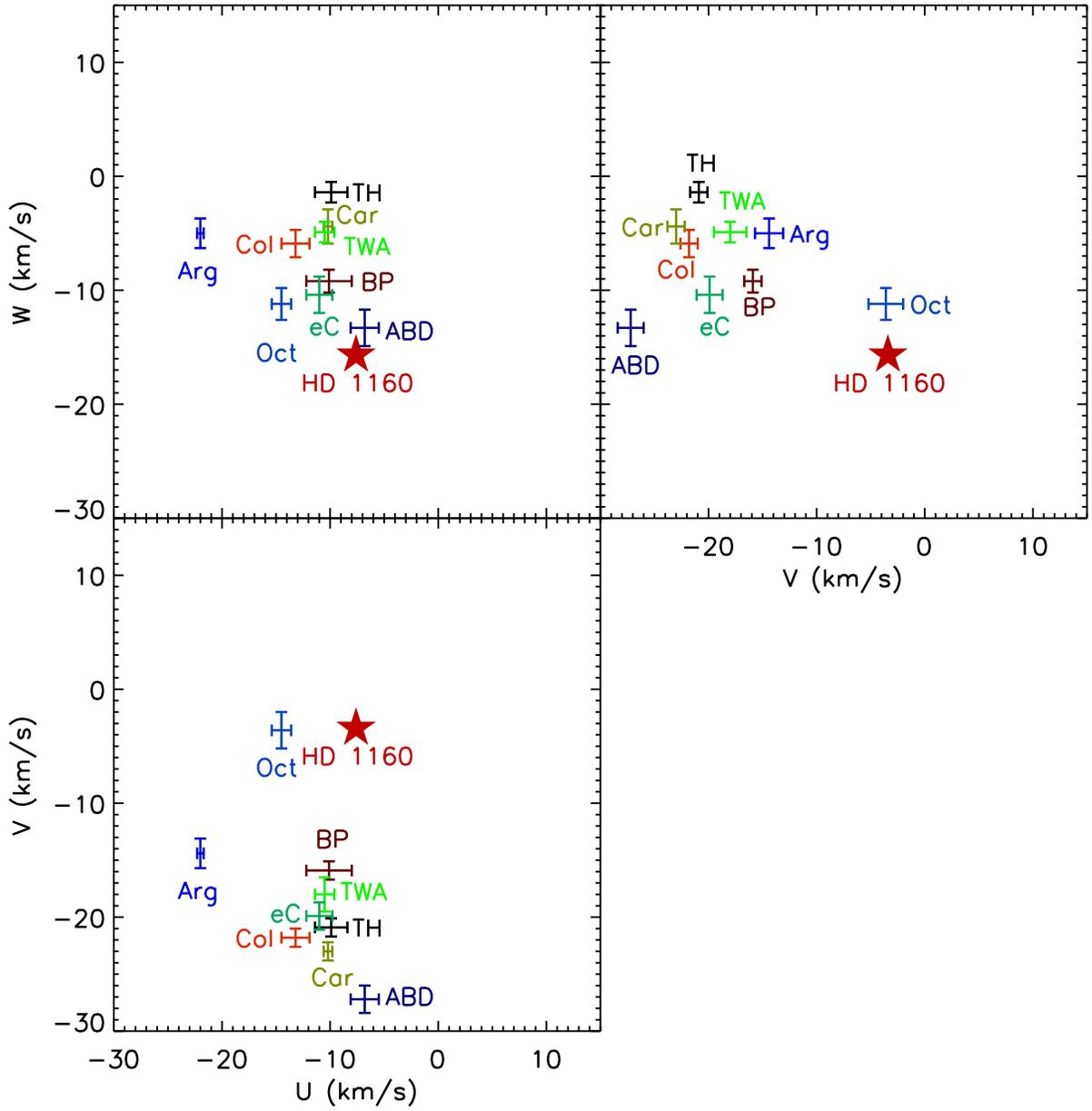}
\caption{UVW motions of nearby young, moving groups from 
\citet{torres08} compared to the HD 
1160 system (BP: $\beta$ Pic, TH: Tuc/Hor, Col: Columba, Car: Carina, 
TWA: TW Hya, eC: $\eta$ Cha, Oct: Octans, Arg: Argus, ABD: AB Dor).  
Error bars for the moving groups indicate the RMS of 
the space motions for the association members, while the 
error bars for HD 1160 indicate measurement uncertainties.  HD 1160 
is not co-moving with any well-known association, though the magnitude 
of its space motion is consistent with these young stars, supporting a 
young age ($<$100 Myr) for HD 1160.
\label{smplot}}
\end{figure}

Near-IR 
spectra of HD~1160~B will likely set much stronger constraints on the age 
of the system, as an $\sim$L0 brown dwarf (as suggested by the photometry 
shown in Figure~\ref{kml_colmag}) will show significant spectral evolution up 
to the main sequence lifetime of HD~1160~A ($\sim$300 Myr).  If the 
spectrum shows signs of low surface gravity compared to field L dwarfs 
(e.g. \citealt{mcgovern04,allers07,allers09}), 
this will help refine our age determination of the system.

In the longer term, high-precision orbital monitoring of both components 
may be able to constrain the 
orbital parameters of HD~1160~B and C.  Given the expected orbital periods of 
$\sim$400 and $\sim$7000 years, such orbital motion will be quite 
small and so will require high precision and long time baselines.  In 
fact, the orbital period of 
HD~1160~B may be significantly longer than 400 years, since 
a circular face-on orbit would predict $\sim$120 mas of motion over 
the 10 years for which we have archival data, and a linear fit to the relative 
motion of HD~1160~B gives only 23 $\pm$ 29 mas over 10 years 
($<$ 1$\sigma$, so not a significant detection of orbital motion).  The 
fact that we observe a larger degree of orbital 
motion from the further-out C (60 $\pm$ 30 mas over 10 years) 
than from the close-in B suggests that there may be 
significant misalignment of the orbital planes of these two 
companions.  
Additional follow-up and analysis of this system will set 
interesting constraints for modeling the formation of HD~1160~ABC.

\acknowledgments

B.A.B was supported by Hubble Fellowship grant HST-HF-01204.01-A awarded by 
the Space Telescope Science Institute, which is operated by AURA for 
NASA, under contract NAS 5-26555.  This work was supported in part by NSF 
grants AST-0713881 and AST-0709484.  
The Gemini Observatory is operated by the Association of Universities for
Research in Astronomy, Inc., under a cooperative agreement with the NSF on
behalf of the Gemini partnership: the National Science Foundation (United
States), the Science and Technology Facilities Council (United Kingdom), the
National Research Council (Canada), CONICYT (Chile), the Australian Research
Council (Australia), CNPq (Brazil), and CONICET (Argentina).  
Based on observations made with the European Southern Observatory
telescopes obtained from the ESO/ST-ECF Science Archive Facility.  This 
publication makes use of data products from the Two Micron All Sky Survey, 
which is a joint project of the University of Massachusetts and the Infrared 
Processing and Analysis Center/California Institute of Technology, funded by 
the National Aeronautics and Space Administration and the National Science 
Foundation.  This research has made use of the SIMBAD database,
operated at CDS, Strasbourg, France.  This research has made use of the 
VizieR catalogue access tool, CDS, Strasbourg, France.  
Some of the data presented herein were 
obtained at the W.M. Keck Observatory, which is operated as a scientific 
partnership among the California Institute of Technology, the University of 
California and the National Aeronautics and Space Administration. The 
Observatory was made possible by the generous financial support of the 
W.M. Keck Foundation.  This paper uses data from the Infrared Telescope 
Facility, which is operated by the University of Hawaii under Cooperative 
Agreement no. NNX-08AE38A with the National Aeronautics and Space 
Administration, Science Mission Directorate, Planetary Astronomy Program.

{\it Facilities:} \facility{Gemini:South (NICI)}, 
\facility{Keck II (NIRC2)},
\facility{VLT:Yepun (NACO)},
\facility{VLT:Melipal (ISAAC)},
\facility{IRTF (SpeX)},
\facility{Magellan II (MIKE)}.

\bibliographystyle{apj}
\bibliography{apj-jour,hd1160}
\clearpage

\clearpage

\begin{deluxetable}{cc}
\tablecaption{Properties of HD~1160~A\label{table1}}
\tablewidth{0pt}
\tablehead{
\colhead{Property} & \colhead{Measurement}
}
\startdata
Spectral Type & A0V \\
RA (ep J2000) & 00:15:57.3025 \\
Dec (ep J2000) & +04:15:04.018 \\
Proper Motion RA (mas/yr) & 21.15 $\pm$ 0.62 \tablenotemark{a} \\
Proper Motion Dec (mas/yr) & $-$14.20 $\pm$ 0.24 \tablenotemark{a} \\
Parallax (mas) & 9.66 $\pm$ 0.45 \tablenotemark{a} \\
Radial Velocity (km/s) & 12.6 $\pm$ 0.3 \tablenotemark{b} \\
$U$ (km/s) & $-$7.6 $\pm$ 0.4 \tablenotemark{b} \\
$V$ (km/s) & $-$3.4 $\pm$ 0.5 \tablenotemark{b} \\
$W$ (km/s) & $-$15.7 $\pm$ 0.4 \tablenotemark{b} \\
$J$ (mag) & 6.98 $\pm$ 0.02 \tablenotemark{c} \\
$H$ (mag) & 7.01 $\pm$ 0.02 \tablenotemark{c} \\
$K_S$ (mag) & 7.04 $\pm$ 0.03 \tablenotemark{c} \\
$L^\prime$ (mag) & 7.055 $\pm$ 0.014 \tablenotemark{d} \\
$M_S$ (mag) & 7.04 $\pm$ 0.02 \tablenotemark{d} \\
Age (Myr) & 50$^{+50}_{-40}$ \tablenotemark{b} \\
\enddata
\tablenotetext{a}{\citet{newhip}}
\tablenotetext{b}{This work}
\tablenotetext{c}{\citet{2mass}}
\tablenotetext{d}{\citet{leggett_keck}}
\end{deluxetable}

\begin{deluxetable}{cccccc}
\tablecaption{Astrometry of HD~1160~BC\label{table2}}
\tablewidth{0pt}
\tablehead{
 & \multicolumn{2}{c}{HD~1160~B} & \multicolumn{2}{c}{HD~1160~C} \\
\colhead{Epoch} & \colhead{Sep (``)} & \colhead{PA ($^{\circ}$)} & \colhead{Sep (``)} & \colhead{PA ($^{\circ}$)} & \colhead{Instrument}
}
\startdata
2002.5699 & 0.77 $\pm$ 0.03 & 246.2 $\pm$ 1.0 & 5.17 $\pm$ 0.03 & 350.9 $\pm$ 0.5 & VLT NACO $K_S$ \\
2002.7397 & \ldots & \ldots & 5.17 $\pm$ 0.04 & 349.8 $\pm$ 0.5 & VLT ISAAC $L$ \\
2002.8658 & \ldots & \ldots & 5.22 $\pm$ 0.04 & 349.6 $\pm$ 0.5 & VLT ISAAC $L$ \\
2003.5342 & \ldots & \ldots & 5.19 $\pm$ 0.04 & 349.5 $\pm$ 0.5 & VLT ISAAC $L$ \\
2003.6356 & \ldots & \ldots & 5.18 $\pm$ 0.04 & 349.7 $\pm$ 0.5 & VLT ISAAC $L$ \\
2003.8384 & 0.77 $\pm$ 0.03 & 245.6 $\pm$ 1.0 & 5.16 $\pm$ 0.03 & 350.1 $\pm$ 0.5 & VLT NACO $K_S$ \\
2003.8603 & \ldots & \ldots & 5.18 $\pm$ 0.04 & 349.6 $\pm$ 0.5 & VLT ISAAC $L$ \\
2005.9753 & 0.76 $\pm$ 0.03 & 244.7 $\pm$ 1.0 & 5.15 $\pm$ 0.03 & 350.4 $\pm$ 0.5 & VLT NACO $L^{\prime}$\\
2007.9014 & \ldots & \ldots & 5.08 $\pm$ 0.07 & 349.4 $\pm$ 0.5 & VLT ISAAC $K_{s}$ \\
2008.5027 & 0.80 $\pm$ 0.06 & 245.3 $\pm$ 2 & 5.16 $\pm$ 0.03 & 349.7 $\pm$ 0.5 & VLT NACO $L^{\prime}$\\
2010.7096 & 0.77 $\pm$ 0.06 & 242.8 $\pm$ 2 & 5.15 $\pm$ 0.03 & 349.4 $\pm$ 0.5 & VLT NACO $L^{\prime}$\\
2010.8301 & 0.78 $\pm$ 0.03 & 244.3 $\pm$ 0.2 & 5.15 $\pm$ 0.03 & 349.8 $\pm$ 0.2 & Gemini NICI $H$\\
2010.8904 & 0.76 $\pm$ 0.03 & 244.5 $\pm$ 0.2 & 5.16 $\pm$ 0.03 & 349.6 $\pm$ 0.2 & Gemini NICI $H$\\
2010.9041 & 0.78 $\pm$ 0.02 & 244.9 $\pm$ 0.5 & 5.14 $\pm$ 0.02 & 349.9 $\pm$ 0.5 & Keck NIRC2 $L^{\prime}$\\
2011.5233 & 0.78 $\pm$ 0.03 & 244.0 $\pm$ 1.0 & 5.14 $\pm$ 0.03 & 349.4 $\pm$ 0.5 & VLT NACO $L^{\prime}$ \\
2011.6685 & 0.78 $\pm$ 0.03 & 244.9 $\pm$ 1.0 & 5.14 $\pm$ 0.03 & 349.5 $\pm$ 0.5 & VLT NACO $L^{\prime}$\\
2011.7589 & \ldots & \ldots & 5.12 $\pm$ 0.07 & 349.2 $\pm$ 0.5 & VLT ISAAC $K_{s}$\\
2011.8027 & 0.77 $\pm$ 0.03 & 244.5 $\pm$ 0.2 & 5.16 $\pm$ 0.03 & 349.6 $\pm$ 0.2 & Gemini NICI $H$\\
2011.8521 & 0.78 $\pm$ 0.03 & 244.4 $\pm$ 1.0 & 5.14 $\pm$ 0.03 & 349.4 $\pm$ 0.5 & VLT NACO $L^{\prime}$\\
\enddata
\end{deluxetable}

\begin{deluxetable}{cccc}
\tablecaption{Photometry of HD~1160~BC\label{table3}}
\tablewidth{0pt}
\tablehead{
\colhead{Bandpass} & \colhead{HD~1160~B} & \colhead{HD~1160~C} & \colhead{Instrument} \\
(MKO) & (mag) & (mag) & 
}
\startdata
$\Delta J$ & 8.85 $\pm$ 0.10 & 6.33 $\pm$ 0.04 & Gemini NICI \\
$\Delta H$ & 7.64 $\pm$ 0.08 & 5.53 $\pm$ 0.03 & Gemini NICI \\
$\Delta K_S$ & 7.08 $\pm$ 0.05 & 5.14 $\pm$ 0.06 & Gemini NICI \\
$\Delta L^{\prime}$ & 6.35 $\pm$ 0.12 & 4.803 $\pm$ 0.005 & Keck NIRC2\\
$\Delta M_S$ & 7.3 $\pm$ 0.2 & 5.10 $\pm$ 0.05 & Keck NIRC2\\
M$_{J}$ & 10.75 $\pm$ 0.14 & 8.23 $\pm$ 0.11 & \\
M$_{H}$ & 9.57 $\pm$ 0.13 & 7.46 $\pm$ 0.11 & \\
M$_{K_S}$ & 9.04 $\pm$ 0.12 & 7.10 $\pm$ 0.12 & \\
M$_{L^{\prime}}$ & 8.33 $\pm$ 0.16 & 6.78 $\pm$ 0.10 & \\
M$_{M_S}$ & 9.3 $\pm$ 0.2 & 7.06 $\pm$ 0.11 & \\
\enddata
\end{deluxetable}

\begin{deluxetable}{ccc}
\tablecaption{Mass Estimates for HD~1160~BC\label{table4}}
\tablewidth{0pt}
\tablehead{
\colhead{Parameter} & \colhead{Measurement} & \colhead{Mass} \\
\colhead{} & \colhead{} & \colhead{(M$_{Jup}$)}
}
\startdata
\multicolumn{3}{c}{HD~1160~B} \\
\hline
M$_{J}$ (mag) & 10.75 $\pm$ 0.14 & 33$^{+12}_{-9}$ \tablenotemark{a}\\
M$_{H}$ (mag) & 9.57 $\pm$ 0.13 & 48$^{+17}_{-13}$  \tablenotemark{a}\\
M$_{K_S}$ (mag) & 9.04 $\pm$ 0.12  & 53$^{+22}_{-15}$  \tablenotemark{a}\\
M$_{L^{\prime}}$ (mag) & 8.3 $\pm$ 0.16 & 59$^{+31}_{-10}$  \tablenotemark{a}\\
M$_{M_S}$ (mag) & 9.3 $\pm$ 0.2 & 33$^{+20}_{-12}$  \tablenotemark{a}\\
BC$_J$  & 2.06 $\pm$ 0.14 \tablenotemark{b} & 37$^{+12}_{-12}$  \tablenotemark{a}\\
\hline
\multicolumn{3}{c}{HD~1160~C} \\
\hline
Sp. Type & M3.5 $\pm$ 0.5 & \\
BC$_{K}$ & 2.72 $\pm$ 0.06 \tablenotemark{c} & \\
\multicolumn{3}{c}{Dwarf Temperature Scale} \\
T$_{eff}$ & 3270 $\pm$ 90 \tablenotemark{d} & 190$^{+65}_{-40}$ \tablenotemark{e} \\
\multicolumn{3}{c}{Intermediate Temperature Scale} \\
T$_{eff}$ & 3340 $\pm$ 70 \tablenotemark{d} & 230$^{+30}_{-45}$ \tablenotemark{e} \\
\enddata
\tablenotetext{a}{\citet{chabrier00}}
\tablenotetext{b}{\citet{liu10}}
\tablenotetext{c}{\citet{sptyperef2}}
\tablenotetext{d}{\citet{luhman99}}
\tablenotetext{e}{\citet{nextgen}}
\end{deluxetable}

\clearpage

\end{document}